\documentclass[aps,prb,showpacs,showkeys,preprintnumbers,amsmath,amssymb,amsfonts,amsfontssuperscriptaddress,twocolumn,nofootinbib]{revtex4}
\usepackage{graphicx}
\usepackage{epstopdf}
\usepackage{color}
\usepackage{accents}
\usepackage{bm}
\usepackage[colorlinks=true,linkcolor=blue,citecolor=red]{hyperref}%
\usepackage{url}
\usepackage[symbol]{footmisc}

\newlength{\dhatheight}

\def\xo{\bm{y}}

\def\s{\bm{s}}
\def\x{{\bm{x}}}
\def\n{\bm{n}}
\def\L{\bm{L}}

\def\nmax{n_{\rm max}}
\def\MM{\mathbf{M}}
\def\A{\mathbf{A}}

\def\F{\mathbf{F}}

\def\I{\mathbf{I}}
\def\J{\mathbf{J}}
\def\O{\mathbf{0}}
\def\U{\mathbf{U}}

\def\W{\mathbf{W}}

\def\PPsi{\bm{\Psi}}

\def\R{\mathbb{R}}
\def\C{\mathbb{C}}

\def\M{\mathcal{M}}
\def\T{\mathcal{T}}

\def\p{\mathbf{p}}
\def\q{\mathbf{q}}

\def\qq{\bm{q}}

\def\pa{{\partial\Omega}}

\begin{document}

\title{Diffusion toward non-overlapping partially reactive spherical traps: \\ fresh insights onto classic problems}

\author{Denis~S.~Grebenkov}
 \email{denis.grebenkov@polytechnique.edu}

\affiliation{Laboratoire de Physique de la Mati\`{e}re Condens\'{e}e (UMR 7643), \\ 
CNRS -- Ecole Polytechnique, IP Paris, 91128 Palaiseau, France}

\date{Received: \today / Revised version: }

\begin{abstract}
Several classic problems for particles diffusing outside an arbitrary
configuration of non-overlapping partially reactive spherical traps in
three dimensions are revisited.  For this purpose, we describe the
generalized method of separation of variables for solving boundary
value problems of the associated modified Helmholtz equation.  In
particular, we derive a semi-analytical solution for the Green
function that is the key ingredient to determine various
diffusion-reaction characteristics such as the survival probability,
the first-passage time distribution, and the reaction rate.  We also
present modifications of the method to determine numerically or
asymptotically the eigenvalues and eigenfunctions of the Laplace
operator and of the Dirichlet-to-Neumann operator in such perforated
domains.  Some potential applications in chemical physics and
biophysics are discussed, including diffusion-controlled reactions for
mortal particles.
\end{abstract}


% 02.50.-r  (Probability theory, stochastic processes, and statistics)
% 05.60.-k  (Transport processes)
% 05.10.-a  (Computational methods in statistical physics and nonlinear dynamics) 
% 02.70.Rr  (General statistical methods)

\maketitle

\section{Introduction}
\label{sec:intro}

Diffusion-reaction processes in industrial chemical reactors, living
cells, and biological tissues have been studied over many decades
\cite{Rice85,Barzykin01,Lauffenburger,Metzler,Lindenberg}.  Diffusion
toward spherical traps (or sinks) is an emblematic model of such
processes that attracted a considerable attention among theoreticians
\cite{Jeffrey73,Kayser83,Kayser84,Felderhof85,Mattern87,Torquato86,Richards87,Rubinstein88,Torquato91,Torquato97,Kansal02}.
In a basic setting, one considers the concentration of diffusing
particles $c(\x,t)$ that obeys diffusion equation in the complement
$\Omega$ of the union of non-overlapping balls:
\begin{equation}  \label{eq:diff}
\frac{\partial}{\partial t} c(\x,t) = D \nabla^2 c(\x,t) ,
\end{equation}
where $D$ is the diffusion coefficient, and $\nabla^2$ is the Laplace
operator.  This equation is completed by an initial concentration
profile, $c(\x,t=0) = c_0(\x)$, an appropriate boundary condition
describing reactions on the boundary $\pa$, and the regularity
condition $c(\x,t) \to 0$ as $|\x| \to \infty$.  Various arrangments
of traps may account for spatial heterogeneities and help to elucidate
the role of disorder onto reaction kinetics, in particular, onto the
reaction rate
\cite{Berezhkovskii90,Berezhkovskii92,Berezhkovskii92b,Makhnovskii93,Oshanin98,Makhnovskii99,Makhnovskii02}.
More generally, reactive traps and passive spherical obstacles can be
used as elementary ``bricks'' to build up model geometrical structures
of porous media or macromolecules such as enzymes or proteins
\cite{Traytak96,Traytak06,Lavrentovich13,Traytak13,Piazza15,Galanti16a,Galanti16b,Grebenkov19}.

As explicit analytical solutions to Eq. (\ref{eq:diff}) are in general
not available, various mathematical tools and numerical techniques
have been broadly used.  For instance, Torquato and co-workers applied
the variational principle to derive upper and lower bounds on the
steady-state reaction rate \cite{Richards87,Rubinstein88,Torquato91}.
Among numerical techniques, Monte Carlo simulations and finite-element
methods were most often employed thanks to their flexibility and
applicability to arbitrary confining domains (see
\cite{Lee89,Tsao01,Eun13,Eun20} and references therein).  In contrast,
the generalized method of separation of variables (GMSV), also known
as the (multipole) re-expansion method, exploits the intrinsic local
symmetries of perforated domains and relies on the re-expansion
(addition) theorems.  This method was applied in different disciplines
ranging from electrostatics to hydrodynamics and scattering theory
\cite{Ivanov70,Martin,Koc98,Gumerov02,Gumerov05}.  In chemical physics,
the GMSV for the Laplace equation was used to study steady-state
diffusion and to compute the reaction rate in various configurations
of traps
\cite{Piazza15,Galanti16a,Galanti16b,Grebenkov19,Goodrich67,Traytak92,Tsao02,McDonald03,Traytak18}.
In particular, a semi-analytical representation for the Green function
of the Laplace equation was derived both in three-dimensional
\cite{Grebenkov19} and two-dimensional spaces \cite{Chen09}, allowing
one to access most steady-state characteristics of the
diffusion-reaction process such as the reaction rate, the escape
probability, the mean first-passage time, the residence time, and the
harmonic measure density.  However, these results are not applicable
to transient time-dependent diffusion among traps, which is governed
by diffusion equation.  As the Laplace transform reduces
Eq. (\ref{eq:diff}) to the modified Helmholtz equation (see below), it
would be natural to adapt the GMSV to this setting.  While the GMSV
for ordinary Helmholtz equation has been broadly employed in
scattering theory \cite{Ivanov70,Martin,Koc98,Gumerov02,Gumerov05},
its applications to the modified Helmholtz equation seem to be much
less studied \cite{Traytak08,Gordeliy09}.

In this paper, we employ re-expansion formulas in spherical domains to
develop a general framework for solving boundary value problems for
the modified Helmholtz equation with Robin boundary conditions
(specified below).  From the numerical point of view, the proposed
method can be seen as an extension of Ref. \cite{Grebenkov19} from the
Laplace equation to the modified Helmholtz equation, as well an
extension of Ref. \cite{Gordeliy09} from exterior to interior domains.
From the theoretical point of view, we derive a semi-analytical
representation of the Green function for the modified Helmholtz
equation which determines most relevant characteristics of transient
time-dependent diffusion.  Moreover, we discuss how this method can be
adapted to compute the eigenvalues and eigenfunctions of the Laplace
operator and of the Dirichlet-to-Neumann operator in such perforated
domains.  To our knowledge, these spectral applications of the method
are new.

The paper is organized as follows.  Section \ref{sec:framework}
presents the GMSV and its applications to get the Green function
(Sec. \ref{sec:Green}), the heat kernel (Sec. \ref{sec:heat}), the
Laplacian spectrum (Sec. \ref{sec:Laplace}) and the spectrum of the
Dirichlet-to-Neumann operator (Sec. \ref{sec:DN}).  In
Sec. \ref{sec:discussion}, we describe practical aspects of these
results and their applications in chemical physics.  In particular, we
discuss first-passage properties (Sec. \ref{sec:first}), stationary
diffusion of mortal particles (Sec. \ref{sec:mortal}), as well as
advantages, limitations and further extensions of the method
(Secs. \ref{sec:advantages}, \ref{sec:extensions}).  Section
\ref{sec:conclusion} concludes the paper.  Appendices regroup
technical derivations and some examples.

\section{General framework}
\label{sec:framework}

\begin{figure}
\begin{center}
\includegraphics[width=80mm]{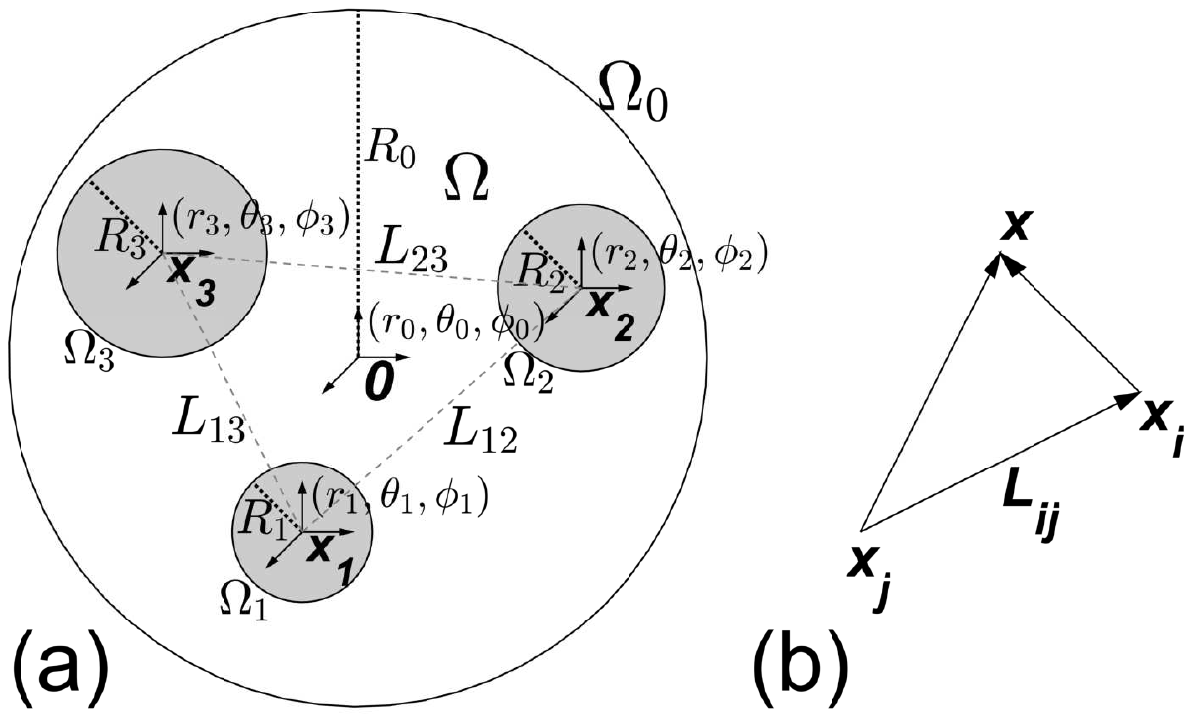}
\end{center}
\caption{
{\bf (a)} Illustration of a bounded perforated domain $\Omega =
\Omega_0 \backslash \bigcup\nolimits_{i=1}^3\overline{\Omega }_i$ with
three balls $\Omega_i$ of radii $R_i$, centered at $\x_i$, all
englobed inside a larger ball $\Omega_0$ of radius $R_0$ centered at
the origin.  Local spherical coordinates, $(r_i,\theta_i,\phi_i)$, are
associated with each ball.  The exterior problem corresponds to the
limit $R_0 = \infty$ when $\Omega_0 = \R^3$.  {\bf (b)} Any point $\x$
can be represented either in local spherical coordinates
$(r_j,\theta_j,\phi_j)$, associated with the center $\x_j$, or in
local spherical coordinates $(r_i,\theta_i,\phi_i)$, associated with
the center $\x_i$.  Accordingly, basis functions
$\psi_{mn}^{\pm}(\x-\x_j)$ can be re-expanded on basis functions
$\psi_{kl}^{\pm}(\x-\x_i)$, where $\x-\x_j = \L_{ij} + (\x-\x_i)$,
with $\L_{ij} = \x_i - \x_j$ being the vector connecting $\x_j$ to
$\x_i$.}
\label{fig:schema}
% A_TraytakG_scheme_fig5a(1);
\end{figure}

We consider diffusion outside the union of $N$ non-overlapping balls
$\Omega_1,\ldots,\Omega_N$ of radii $R_i$, centered at $\x_i$:
\begin{equation}
\Omega = \Omega_0 \backslash \bigcup\limits_{i=1}^N \overline{\Omega}_i ,  \quad \Omega_i = \{ \x\in\R^3 ~:~ |\x - \x_i|<R_i\},
\end{equation}
where $\Omega_0$ is a ball of radius $R_0$, centered at the origin
$\x_0 = 0$, that englobes all the balls: $\overline{\Omega}_i \subset
\Omega_0$ for all $i$ (Fig. \ref{fig:schema}).  We allow $R_0$ to be
infinite (i.e., $\Omega_0 = \R^3$) that describes an exterior problem,
in which particles diffuse in an unbounded domain $\Omega$ and thus
can escape at infinity.  In turn, for any finite $R_0$, one deals with
an interior problem of diffusion in a bounded domain $\Omega$.

\subsection{General boundary value problem}
\label{sec:general}

We first consider a general boundary value problem for the modified
Helmholtz equation
\begin{subequations}  \label{eq:Helm_problem}
\begin{eqnarray}  \label{eq:Helm}
(q^2 - \nabla^2) w(\x;q) &=& 0  \quad (\x\in \Omega), \\  \label{eq:Robin}
\left.\left(a_i w + b_i R_i \frac{\partial w}{\partial \n}\right)\right|_{\pa_i} &=& f_i \quad (i = 0,\ldots,N),
\end{eqnarray}
\end{subequations}
where $q$ is a nonnegative parameter%
\footnote{
While we focus on nonnegative $q$ throughout the main text, the method
is implemented for any complex $q$, see Appendix \ref{sec:complex_q}.},
%footnote
$\partial/\partial \n$ is the normal derivative on the boundary $\pa =
\cup_{i=0}^N \pa_i$, oriented outwards the domain $\Omega$, $f_i$ are
given continuous functions on $\pa_i$, and $a_i$ and $b_i$ are
nonnegative constants such that $a_i + b_i > 0$ (i.e., $a_i$ and $b_i$
cannot be simultaneously $0$).  The Robin boundary condition
(\ref{eq:Robin}) is reduced to Dirichlet condition for $b_i = 0$ and
to Neumann condition for $a_i = 0$.  In particular, our description
can accommodate perfectly reactive traps or sinks ($a_i > 0$, $b_i =
0$), partially reactive traps ($a_i > 0$, $b_i > 0$), and passive
reflecting obstacles ($a_i = 0$, $b_i > 0$).  For the exterior
problem, Eq. (\ref{eq:Robin}) for $i = 0$ is replaced by the
regularity condition $w(\x;q) \to 0$ as $|\x|\to\infty$.

The basic idea of the GMSV consists in searching for the solution of
Eq. (\ref{eq:Helm}) as a superposition of partial solutions $w_i$ in
the exterior of each ball $\Omega_1,\ldots,\Omega_N$, and in the
interior of $\Omega_0$:
\begin{equation}  \label{eq:g_gi}
w(\x;q) = \sum\limits_{i=0}^N w_i(\x;q) 
\end{equation}
(for the exterior problem, $w_0 \equiv 0$).  As each domain $\Omega_i$
is spherical, the corresponding partial solution can be searched in
the {\it local} spherical coordinates $(r_i,\theta_i,\phi_i)$
associated with $\Omega_i$, as an expansion over regular (for $i = 0$)
and irregular (for $i > 0$) basis functions $\psi_{mn}^{\pm}$ with
unknown coefficients $A_{mn}^i$,
\begin{equation}  \label{eq:gi}
w_i(\x;q) = \sum\limits_{n=0}^\infty \sum\limits_{m=-n}^n A_{mn}^i \, \psi_{mn}^{\epsilon_i}(q r_i,\theta_i,\phi_i),
\end{equation}
where we use a shortcut notation $\epsilon_i = -$ for $i > 0$, and
$\epsilon_0 = +$.  For the modified Helmholtz equation, the basis
functions are
\begin{equation}
\begin{split}
\psi_{mn}^{+}(qr_i,\theta_i,\phi_i) &= i_n(qr_i) \, Y_{mn}(\theta_i,\phi_i) , \\
\psi_{mn}^{-}(qr_i,\theta_i,\phi_i) &= k_n(qr_i) \, Y_{mn}(\theta_i,\phi_i) , \\
\end{split}
\end{equation}
where
\begin{equation}
\begin{split}
i_n(z) & = \sqrt{\pi/(2z)} \, I_{n+1/2}(z), \\
k_n(z) & = \sqrt{2/(\pi z)} \, K_{n+1/2}(z) \\
\end{split}
\end{equation}
are the modified spherical Bessel functions of the first and second
kind, and $Y_{mn}(\theta,\phi)$ are the normalized spherical
harmonics:
\begin{equation}  \label{eq:Y}
Y_{mn}(\theta,\phi) = \sqrt{\frac{(2n+1) \, (n-m)!}{4\pi \, (n+m)!}} \, P_n^m(\cos\theta) \, e^{im\phi} ,
\end{equation}
with $P_n^m(z)$ being the associated Legendre polynomials (we use the
convention that $Y_{mn}(\theta,\phi) \equiv 0$ for $|m| > n$).

The unknown coefficients $A_{mn}^i$ are fixed by the boundary
condition (\ref{eq:Robin}) applied on each $\pa_i$:
\begin{equation} \label{eq:fi_def}
f_i = \sum\limits_{j=0}^N \sum\limits_{m,n} A_{mn}^j
 \biggl(a_i + b_i R_i \frac{\partial}{\partial \n}\biggr) \psi_{mn}^{\epsilon_j}(qr_j,\theta_j,\phi_j) \biggr|_{\pa_i} ,
\end{equation}
where $\sum\nolimits_{m,n}$ is a shortcut notation for the sum over $n
= 0,1,2,\ldots$ and $m = -n,-n+1,\ldots,n$.  As spherical harmonics
form a complete basis of the space $L_2(\pa_i)$, one can project this
functional equation onto $Y_{kl}(\theta_i,\phi_i)$ to reduce it to an
infinite system of linear algebraic equations on the coefficients
$A_{mn}^j$:
\begin{equation}  \label{eq:coeff}
F_{kl}^i =  \sum\limits_{j=0}^N \sum\limits_{m,n} A_{mn}^j \, W_{mn,kl}^{j,i}  
\quad \begin{cases}  i=0,1,\ldots,N, \\  l=0,1,\ldots,\,  |k|\leq l, \end{cases}
\end{equation}
where
\begin{align}   \label{eq:W_def}
& W_{mn,kl}^{j,i} \\  \nonumber
& = \biggl( \biggl(a_i + b_i R_i \frac{\partial}{\partial \n}\biggr) \psi_{mn}^{\epsilon_j}(qr_j,\theta_j,\phi_j) \biggr|_{\pa_i} , 
Y_{kl} \biggr)_{L_2(\pa_i)}  
\end{align}
and
\begin{equation}  \label{eq:F_def}
F_{kl}^i = (f_i, Y_{kl})_{L_2(\pa_i)} ,
\end{equation}
with the standard scalar product: $(f,g)_{L_2(\pa_i)} =
\int\nolimits_{\pa_i} d\s \, f(\s) \, g^*(\s)$, asterisk denoting the
complex conjugate.  Even though $A_{mn}^j$, $F_{mn}^i$, and
$W_{mn,kl}^{j,i}$ involve many indices, one can re-order them to
consider $A_{mn}^j$ (resp., $F_{mn}^i$) as components of a (row)
vector $\A$ (resp., $\F$), while $W_{mn,kl}^{j,i}$ as components of a
matrix $\W$, so that Eq. (\ref{eq:coeff}) becomes a matrix equation:
\begin{equation}  \label{eq:F_AW}
\F = \A \W .
\end{equation}
In Appendix \ref{sec:AW}, we provide the explicit formulas for the
matrix elements $W_{mn,kl}^{j,i}$, which depend only on $q$, on the
positions and radii of the balls $\Omega_i$, and on the parameters
$a_i$ and $b_i$.  The derivation of these formulas relies on the
re-expansion (addition) theorems for basis solutions $\psi_{mn}^{\pm}$
\cite{Hobson,Epton95}.  Truncating the infinite-dimensional matrix
$\W$ and inverting it numerically yield a truncated set of
coefficients $A_{mn}^j$.  In this way, Eqs. (\ref{eq:g_gi},
\ref{eq:gi}) provide a semi-analytical solution of the boundary value
problem (\ref{eq:Helm}, \ref{eq:Robin}), in which the dependence on
$\x$ is analytical (via explicit basis functions $\psi_{mn}^{\pm}$),
but the coefficients $A_{mn}^i$ have to be obtained numerically from
Eq. (\ref{eq:F_AW}).  A practical implementation of this method is
summarized in Appendix \ref{sec:implementation}, whereas its
advantages and limitations are discussed in Sec. \ref{sec:advantages}.

\begin{figure}
\begin{center}
\includegraphics[width=88mm]{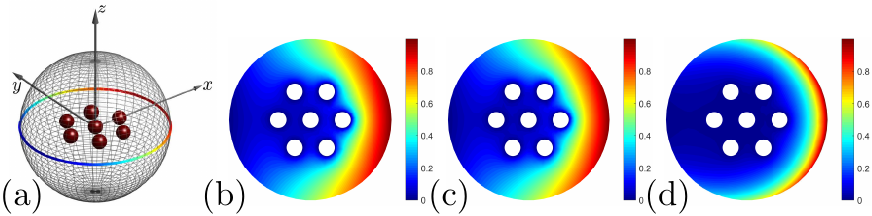}
\end{center}
\caption{
{\bf (a)} Configuration of 7 perfect traps of radius $R_i = 0.1$
inside a larger sphere of radius $R_0 = 1$ on which the variable
concentration profile is set: $f_0(\theta,\phi) = \frac12 (1 +
\sin\theta \cos \phi)$ (illustrated by a colored contour at the
equator).  {\bf (b,c,d)} The solution $w(\x;q)$ evaluated on a
horizontal cut at $z = 0$ (i.e., in the plane $xy$, view from the
top), with $q = 0.2$ {\bf (b)}, $q = 1$ {\bf (c)} and $q = 5$ {\bf
(d)}.  The matrix $\W$ determining the coefficients $A_{mn}^i$ was
truncated to the size $8(3+1)^2 \times 8(3+1)^2 = 128 \times 128$ with
the truncation order $\nmax = 3$. }
\label{fig:conc}
% [A, S, wx,X1] = A_TraytakP_conc_example;
% export_fig('....eps', '-eps');
\end{figure}

Figure \ref{fig:conc} illustrates three solutions $w(\x;q)$ of the
modified Helmholtz equation with Dirichlet boundary conditions on a
configuration with 7 balls enclosed by a larger sphere.  As $q$
increases, the solution $w(\x;q)$ drops faster from its larger values
on the outer sphere toward the perfectly absorbing traps.

\subsection{Green function}
\label{sec:Green}

The above general solution allows one to derive many useful
quantities.  Here, we aim at finding the Green function $G(\x,\xo;q)$
of the modified Helmholtz equation in $\Omega$ \cite{Duffy,Keilson}
\begin{subequations}
\begin{eqnarray}  \label{eq:Helm2}
(q^2 - \nabla^2) G(\x,\xo;q) &=& \delta(\x-\xo)  \quad (\x\in \Omega), \\  \label{eq:Robin2}
\left.\left(a_i G + b_i R_i \frac{\partial G}{\partial \n}\right)\right|_{\pa_i} &=& 0 \quad (i = 0,\ldots,N),
\end{eqnarray}
\end{subequations}
where $\delta(\x-\xo)$ is the Dirac distribution, and $\xo$ is a fixed
point in $\Omega$ (for the exterior problem, Eq. (\ref{eq:Robin2}) for
$i=0$ is replaced by regularity condition $G(\x,\xo;q)\to 0$ as
$|\x|\to\infty$).  We search for the Green function in the form
\begin{equation}  \label{eq:Green0}
G(\x,\xo;q) = G_{\rm f}(\x,\xo;q) - g(\x;\xo,q),
\end{equation}
where
\begin{equation}
\label{eq:G0_Helm}
G_{\rm f}(\x,\xo;q) = \frac{\exp(-q|\x - \xo|)}{4\pi |\x-\xo|}
\end{equation}
is the fundamental solution of the modified Helmholtz equation,
whereas the auxiliary function $g(\x;\xo,q)$ satisfies
Eqs. (\ref{eq:Helm_problem}), with
\begin{equation}
f_i = \left.\left(a_i G_{\rm f} + b_i R_i \frac{\partial G_{\rm f}}{\partial \n}\right)\right|_{\pa_i} \,.
\end{equation}
In Appendix \ref{sec:AF}, we derive explicit formulas for the scalar
product in Eq. (\ref{eq:F_def}) determining the components $F_{mn}^i$
of the vector $\F$.

Among various applications, the Green function allows one to solve the
inhomogeneous modified Helmholtz equation:
\begin{subequations}  \label{eq:Helm_inhom}
\begin{eqnarray}  
(q^2 - \nabla^2) w(\x;q) &=& F(\x)  \quad (\x\in \Omega), \\ 
\left.\left(a_i w + b_i R_i \frac{\partial w}{\partial \n}\right)\right|_{\pa_i} &=& 0 \quad (i = 0,\ldots,N)
\end{eqnarray}
\end{subequations}
(with a given continuous function $F$) as
\begin{equation}
w(\x) = \int\limits_{\Omega} d\xo \, G(\x,\xo;q) \, F(\xo).
\end{equation}
Equivalently, Eqs. (\ref{eq:Helm_inhom}) could be solved by reduction
to homogeneous Eqs. (\ref{eq:Helm_problem}) with the help of the
fundamental solution $G_{\rm f}(\x,\xo;q)$.

\subsection{Heat kernel}
\label{sec:heat}

The solution of the modified Helmholtz equation opens a way to
numerous applications in heat transfer and nonstationary diffusion.
For instance, the Green function $G(\x,\xo;q)$ is related to the
Laplace transform of the heat kernel $P(\x,t|\xo)$ that satisfies the
diffusion equation
\begin{subequations}
\begin{eqnarray}
\frac{\partial P(\x,t|\xo)}{\partial t} - D \nabla^2  P(\x,t|\xo) &=& 0  , \\
P(\x,t=0|\xo) &=& \delta(\x-\xo) , \\
\left.\left(a_i P + b_i R_i \frac{\partial P}{\partial \n}\right)\right|_{\pa_i} &=& 0 
\end{eqnarray}
\end{subequations}
(for the exterior problem, the Robin boundary condition on $\pa_0$ is
replaced by the regularity condition $P\to 0$ as $|\x|\to \infty$).
The heat kernel describes the likelihood of the event that a particle
that started from a point $\xo$ at time $0$, is survived against
surface reactions on $\pa$ and found in a vicinity of a point $\x$ at
a later time $t$ \cite{Gardiner,Grebenkov19e}.  The Laplace transform
of the diffusion equation yields the modified Helmholtz equation, so
that
\begin{equation}  \label{eq:P_G}
\int\limits_0^\infty dt \, e^{-pt} \, P(\x,t|\xo) = \frac{1}{D} \, G(\x,\xo; \sqrt{p/D}).
\end{equation}

\subsection{Laplacian eigenvalues and eigenfunctions}
\label{sec:Laplace}

Replacing $q$ by $iq$ transforms the modified Helmholtz equation
(\ref{eq:Helm}) to the ordinary Helmholtz equation: 
\begin{equation} \label{eq:Helm_ordinary}
(q^2 + \nabla^2) w(\x;q) = 0.  
\end{equation}
As solutions of this equation by the GMSV were thoroughly studied in
scattering theory
\cite{Ivanov70,Martin,Koc98,Gumerov02,Gumerov05}, we do not discuss
them here.  However, we mention that the above method can also be
adapted to compute the eigenvalues and eigenfunctions of the Laplace
operator $-\nabla^2$ in a bounded domain $\Omega$ (i.e., with $R_0 <
\infty$):
\begin{subequations}
\begin{eqnarray}
\label{eq:eigen}
\nabla^2 u(\x) + \lambda u(\x) &=& 0 \quad (\x\in \Omega), \\
\left. \left(a_i u + b_i R_i \frac{\partial u}{\partial \n}\right) \right|_{\pa_i} &=& 0 .
\end{eqnarray}
\end{subequations}
As Eq. (\ref{eq:eigen}) is the ordinary Helmholtz equation, it is
convenient to search for an eigenpair $(\lambda, u(\x))$ in the form
\begin{equation}
\label{eq:eigen_u}
u(\x) = \sum\limits_{j=0}^{N} \sum\limits_{m,n} A_{mn}^j\, \psi_{mn}^{\epsilon_j}(qr_j, \theta_j, \phi_j) ,
\end{equation}
with $q = i\sqrt{\lambda}$.  This is equivalent to setting $f_i \equiv
0$ and thus $\F \equiv 0$ in Eq. (\ref{eq:F_AW}).  The necessary and
sufficient condition to satisfy the matrix equation $\A \W = 0$ is
\begin{equation}
\label{eq:eigen_det}
\det(\W) = 0 .
\end{equation}
If $\{q_k\}$ is the set of the values of $q$ at which this condition
is satisfied, one gets the eigenvalues: $\lambda_k = - q_k^2$.  From
the general spectral theory, the Laplace operator in a bounded domain
with Robin boundary conditions is known to have an infinitely many
nonnegative eigenvalues growing to infinity so that all zeros $q_k$
should lie on the imaginary axis.  In practice, the matrix $\W$ is
first truncated and then some zeros $q_k$ of $\det(\W)$ are computed
numerically.  These zeros yield the approximate eigenvalues.

The computation of the associated eigenfunctions is standard.  At each
value $q_k$, the system of linear equations $\A \W = 0$ is
under-determined and has infinitely many solutions.  If the eigenvalue
$\lambda_k = -q_k^2$ is simple, one can fix a solution by setting one
of unknown coefficients, e.g., $A_{00}^1$, to a constant $c$.  This
results in a smaller system of {\it inhomogeneous} linear equations on
the remaining coefficients $A_{mn}^j$ that can be solved numerically.
The corresponding eigenfunction is given by Eq. (\ref{eq:eigen_u}).
The arbitrary constant $c$ is simply a choice of the normalization of
that eigenfunction.  Once the eigenfunction is constructed, it can be
renormalized appropriately.  For eigenvalues with multiplicity $m >
1$, an eigenfunction is defined up to $m$ free constants that can be
chosen in a standard way.

\subsection{Dirichlet-to-Neumann operator}
\label{sec:DN}

The GMSV can be applied to investigate the spectral properties of the
Dirichlet-to-Neumann operator.  For a given function $f$ on the
boundary $\pa$, the Dirichlet-to-Neumann operator $\M_p$ associates
another function $g = (\partial w/\partial \n)_{|\pa}$ on that
boundary, where $w$ is the solution of the Dirichlet boundary value
problem
\begin{equation}  \label{eq:Helm3}
(p - D\nabla^2) w = 0 \quad (\x\in \Omega), \qquad w|_{\pa} = f 
\end{equation}
(for an exterior problem, the regularity condition $w(\x)\to 0$ as
$|\x|\to\infty$ is also imposed; see
\cite{Arendt14,Daners14,Arendt15,Hassell17,Girouard17} for a rigorous
mathematical definition).  The Dirichlet-to-Neumann operator can be
used as an alternative to the Laplace operator in describing
diffusion-reaction processes.  In particular, the eigenvalues and
eigenfunctions of $\M_p$ determine most diffusion-reaction
characteristics, even for inhomogeneous surface reactivity
\cite{Grebenkov19b,Grebenkov19c}.

As the boundary $\pa$ is the union of non-intersecting spheres
$\pa_i$, a function $f$ on $\pa$ can be represented by its
restrictions $f_i = f|_{\pa_i}$, and Eq. (\ref{eq:g_gi}) is the
semi-analytical solution of Eq. (\ref{eq:Helm3}), by setting $q =
\sqrt{p/D}$, $a_i = 1$ and $b_i = 0$.  The action of the operator
$\M_p$ can be determined by computing the normal derivative of the
solution $w$.  In Appendix \ref{sec:dwdn}, we represented the normal
derivative as
\begin{equation} \label{eq:dwdn}
\left. \biggl(\frac{\partial w}{\partial \n}\biggr)\right|_{\pa_i} =  
\sum\limits_{m,n} \bigl(\F \tilde{\W}^{-1} \tilde{\W}'\bigr)_{mn}^i  Y_{mn}(\theta_i,\phi_i) ,
\end{equation}
where the matrices $\tilde{\W}$ and $\tilde{\W}'$ are defined by
explicit formulas (\ref{eq:Wtilde0}, \ref{eq:Wtilde}), and we inverted
Eq. (\ref{eq:F_AW}) to express the coefficients $\A$.  As a
consequence, the Dirichlet-to-Neumann operator $\M_p$ is represented
in the basis of spherical harmonics by the following matrix
\begin{equation}  \label{eq:M_DtN}
\MM = \tilde{\W}^{-1} \tilde{\W}' .
\end{equation}
In particular, the eigenvalues of this matrix coincide with the
eigenvalues of $\M_p$, whereas its eigenvectors allow one to
reconstruct the eigenfunctions of $\M_p$.  In practice, one computes a
truncated version of the matrix $\MM$ so that its eigenvalues would
approximate a number of eigenvalues of $\M_p$.  For the exterior
problem, one needs to reduce the matrices $\tilde{\W}$ and
$\tilde{\W}'$ by removing the block row and block column corresponding
to $\Omega_0$ (see Appendix \ref{sec:AW}).  We recall that, in
contrast to the Laplace operator, whose spectrum is continuous for the
exterior problem, the spectrum of the Dirichlet-to-Neumann operator is
discrete for interior and exterior perforated domains, because their
boundary $\pa$ is bounded in both cases.

The above method can also be adapted to study an extension of the
Dirichlet-to-Neumann to the case when some spheres $\Omega_i$ are
reflecting.  In fact, let $I$ denote the set of indices of spheres
$\pa_i$ that are reactive, whereas the remaining spheres with indices
$\{0,1,\ldots,N\} \backslash I$ are reflecting.  Then one can define
the Dirichlet-to-Neumann operator $\M_p^\Gamma$, acting on a function
$f$ on $\Gamma = \cup_{i\in I} \pa_i$ as $\M_p^\Gamma ~:~ f\to g =
(\partial w/\partial \n)|_{\Gamma}$, where $w$ is the solution of the
mixed boundary value problem:
\begin{equation}
(p - D \nabla^2) w = 0 \quad \textrm{in} ~\Omega, \qquad 
\left\{ \begin{array}{l} w|_{\Gamma} = f , \\ (\partial w/\partial \n)|_{\Omega\backslash \Gamma} = 0.  \end{array} \right.
\end{equation}
The matrix representation of the operator $\M_p^\Gamma$ is still given
by Eq. (\ref{eq:M_DtN}), in which the matrix $\tilde{\W}$ is replaced
by another matrix evaluated with $a_i = 1$, $b_i = 0$ for $i\in I$
(Dirichlet condition) and $a_i = 0$, $b_i = 1$ for $i \in
\{0,1,\ldots,N\} \backslash I$ (Neumann condition), see Appendix
\ref{sec:AW}.

\section{Discussion}
\label{sec:discussion}

The previous section presented a concise overview of several major
applications of the GMSV for the modified Helmholtz equation.  In this
section, we discuss its practical aspects and illustrate the use of
the GMSV on several examples in the context of chemical physics.

\subsection{First-passage properties}
\label{sec:first}

As the Green function $G(\x,\xo;q)$ is related via Eq. (\ref{eq:P_G})
to the Laplace transform of the heat kernel, it determines most
diffusion-reaction characteristics in the Laplace domain (see
\cite{Grebenkov19b} for details).  For instance, the Laplace transform
of the probability flux density $j(\s,t|\xo)$ reads
\begin{align}  \nonumber
\tilde{j}(\s,p|\xo) & = \int\limits_0^\infty dt \, e^{-pt} \, j(\s,t|\xo) \\  \label{eq:jdef}
& = \biggl(-\frac{\partial G(\x,\xo; \sqrt{p/D})}{\partial \n}\biggr)\biggr|_{\x = \s \in \pa} .
\end{align}
We recall that $j(\s,t|\xo)$ is the joint probability density of the
reaction time and location on the partially reactive boundary $\pa$
for a particle started from a point $\xo \in \Omega$.  The normal
derivative of the Green function was evaluated in Appendix
\ref{sec:dwdn}, yielding:
\begin{equation}   \label{eq:tildej}
\tilde{j}(\s,p|\xo)\biggr|_{\pa_i}  = \sum\limits_{m,n} \J_{mn}^i(\xo) \, Y_{mn}(\theta_i,\phi_i)  ,
\end{equation}
where the components of the vector $\J$ are given by
Eq. (\ref{eq:Jmatrix}), with $q = \sqrt{p/D}$.

\subsubsection*{Probability distribution of the reaction time}

The integral of the joint probability density $j(\s,t|\xo)$ over time
$t$ yields the spread harmonic measure density on the sphere $\pa_i$
\cite{Grebenkov06,Grebenkov15}.  This is a natural extension of the
harmonic measure density to partially reactive traps with Robin
boundary condition, which characterizes the distribution of the
reaction location.  As the integral of $j(\s,t|\xo)$ over $t$ is equal
to $\tilde{j}(\s,0|\xo)$ (i.e., with $p = q = 0$), the modified
Helmholtz equation is reduced to the Laplace equation.  The explicit
representation of the spread harmonic measure density and its
properties were discussed in Ref. \cite{Grebenkov19}.  

In turn, the integral of $j(\s,t|\xo)$ over the location position $\s$
yields the probability density of the reaction time:
\begin{equation}
H(t|\xo) = \int\limits_{\pa} d\s \, j(\s,t|\xo) .
\end{equation}
In the Laplace domain, the expansion (\ref{eq:tildej}) allows one to
easily compute this integral due to the orthogonality of spherical
harmonics:
\begin{equation}  \label{eq:tildeH}
\tilde{H}(p|\xo) = \int\limits_{\pa} d\s \, \tilde{j}(\s,p|\xo) = \sqrt{4\pi} \sum\limits_{i=0}^N R_i^2 \, \J_{00}^i(\xo) ,
\end{equation}
where the factor $R_i^2$ accounts for the area of the $i$-th ball, and
the matrix elements $\J_{00}^i(\xo)$ are given in Eq. (\ref{eq:J00i}).
Note that each term in this sum is the probability flux onto the
sphere $\pa_i$, while the dependence on $\xo$ comes explicitly through
the expression for $\J$.  By definition, $\tilde{H}(p|\xo) = \langle
\exp(-p \T)\rangle$ is the generating function of the moments of the
reaction time $\T$:
\begin{equation}
\langle \T^k \rangle = (-1)^k \lim\limits_{p\to 0} \frac{\partial^k \tilde{H}(p|\xo)}{\partial p^k} \,.
\end{equation}
One can thus determine the mean and higher-order moments of the
reaction time $\mathcal T$.  In turn, the inverse Laplace transform of
Eq. (\ref{eq:tildeH}) gives $H(t|\xo)$ in time domain.  The integral
of $H(t|\xo)$ from $0$ to $t$ yields the probability of reaction up to
time $t$, whereas the integral from $t$ to infinity is the survival
probability of the particle.
We conclude that the present approach opens new opportunities for
studying various first-passage phenomena for an arbitrary
configuration of non-overlapping partially reactive spherical traps.
In other words, this approach generalizes the classical results for
diffusion outside a single trap, for which one has $\U = 0$, and the
above expression simplifies to
\begin{equation}   \label{eq:Hp_sphere}
\tilde{H}(p|\xo) = \frac{R_1 \, e^{-\sqrt{p/D}(|\xo|-R_1)}}{|\xo| (a_1 + b_1 (1+R_1\sqrt{p/D}))} \,,
\end{equation}
where we used the Wronskian 
\begin{equation} \label{eq:Wronskian}
i'_n(z) k_n(z) - k'_n(z) i_n(z) = \frac{1}{z^2} \,
\end{equation}
and the explicit relations $i_0(z) = \sinh(z)/z$ and $k_0(z) =
e^{-z}/z$.  The inverse Laplace transform of this formula yields the
expression for $H(t|\xo)$ derived by Collins and Kimball
\cite{Collins49}.  Setting $a_1 = 1$ and $b_1 = 0$, one retrieves
another classical expression for a perfectly reactive trap studied by
von Smoluchowski \cite{Smoluchowski17}.  We emphasize that for a
single trap, the analysis can be pushed much further by including,
e.g., the interaction potentials (see \cite{Sano79,Son13,Lee20} and
references therein).  The more elaborate example of two concentric
spheres is discussed in Appendix \ref{sec:concentric}.

\subsubsection*{Presence of reflecting obstacles?}

\begin{figure}
\begin{center}
\includegraphics[width=88mm]{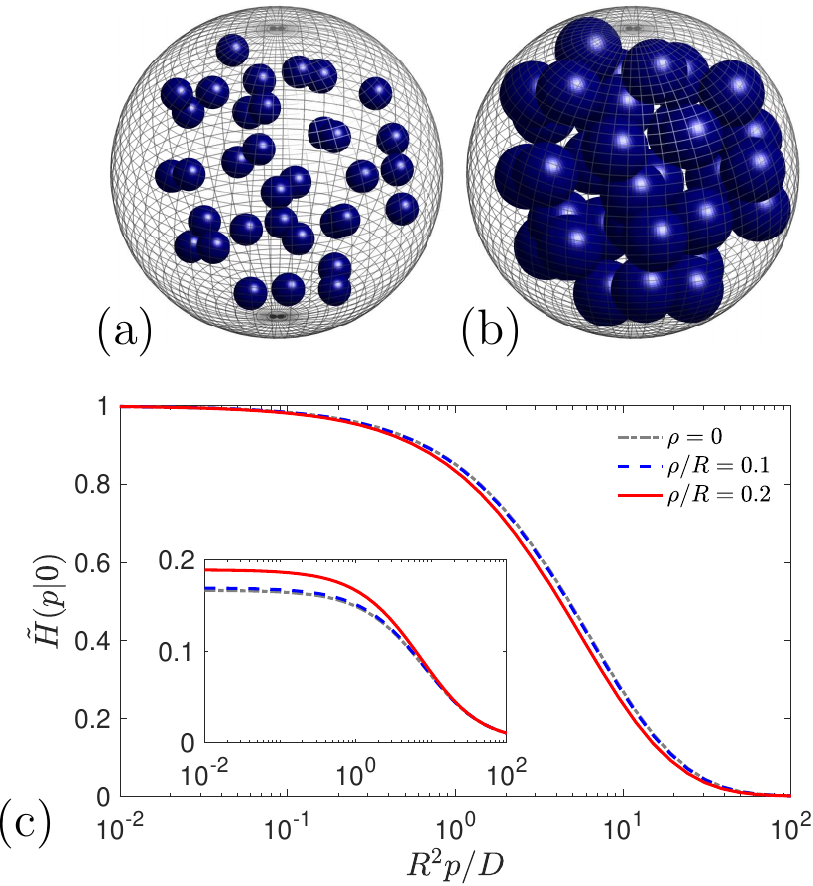}
\end{center}
\caption{
{\bf (a,b)} Two configurations of 35 reflecting spherical obstacles of
radius $\rho$ inside a larger sphere of radius $R$, with the same
centers but distinct radii: $\rho/R = 0.1$ {\bf (a)} and $\rho/R =
0.2$ {\bf (b)}.  {\bf (c)} Laplace-transformed probability density
$\tilde{H}(p|0)$ of the first-exit time from the center of the ball of
radius $R$ to its boundary $\pa_0$ in presence of 35 reflecting
spherical obstacles.  The function $\tilde{H}(p|0)$ was computed via
Eq. (\ref{eq:tildeH}), with the truncation order $\nmax = 2$.  For
comparison, the gray dash-dotted line shows the classical expression
$\tilde{H}(p|0) = 1/i_0(R\sqrt{p/D})$ for an empty ball without
obstacles.  The inset shows the Laplace-transformed survival
probability $\tilde{S}(p|0) = (1 - \tilde{H}(p|0))/p$. }
\label{fig:Hp_origin}
% A_TraytakP_obstacles_exit2_fig(Hp1,Hp2);
% A_TraytakP_show_S(S);
% export_fig('balls35_01.eps', '-eps');
\end{figure}

How do reflecting obstacles modify the reaction time distribution?
Figure \ref{fig:Hp_origin} presents the Laplace-transformed
probability density $\tilde{H}(p|0)$ of the first-exit time from the
center of the ball of radius $R_0 = R$ to its boundary $\pa_0$ in
presence of 35 reflecting spherical obstacles of equal radii $R_i =
\rho$.  In Ref. \cite{Grebenkov17d}, we conjectured that reflecting
obstacles cannot speed up the exit from the center of the ball, i.e.,
$S(t|0) \geq S_0(t|0)$, where $S(t|0)$ and $S_0(t|0)$ are the survival
probabilities with and without obstacles, respectively.  As a
consequence, their Laplace transforms satisfies the same inequality:
$\tilde{S}(p|0) \geq \tilde{S}_0(p|0)$.  This statement is not
trivial: on one hand, reflecting obstacles hinder the motion of the
diffusing particle and thus increase its first-exit time; on the other
hand, the obstacles reduce the available space that might speed up the
exit.  According to this conjecture, the hindering effect always
``wins'' for diffusion from the center to the boundary of a ball, but
it is not necessarily true neither for other starting points, nor for
other (non-spherical) domains.  This conjecture is confirmed in our
numerical example, as illustrated in the inset of
Fig. \ref{fig:Hp_origin}.  Expectedly, small obstacles ($\rho/R =
0.1$) almost do not alter $\tilde{H}(p|0)$ and $\tilde{S}(p|0)$, the
curves being barely distinguishable.  Most surprisingly, even large
obstacles ($\rho/R = 0.2$) that fill $35(\rho/R)^3 \approx 28\%$ of
the volume, also have a very moderate effect, which is mainly visible
on the inset at small $p$.  Indeed, the obstacles hinder diffusion and
slightly increase the mean first-exit time $\tilde{S}(p=0|0)$, from
$R^2/(6D) \approx 0.17 (R^2/D)$ without obstacles, to $0.19 (R^2/D)$
in the presence of obstacles.  Even though this observation is
realized for the particular geometric setting of spherical obstacles,
one can question the role of hindering obstacles in more general
configurations.  A systematic study of this problem can be performed
in future by using the present numerical and analytical approach.
As discussed in Sec. \ref{sec:mortal}, $\tilde{H}(p|0)$ can
alternatively be interpreted as the stationary concentration at $\xo =
0$ of mortal particles whose concentration on the outer boundary is
kept constant.

\subsubsection*{Presence of absorbing sinks?}

With the help of the GMSV, one can refine the above analysis by
considering the following first-passage time problem: for a particle
started from $\xo$, what is the reaction time on a given trap $i$ in
the presence of absorbing sinks that can irreversibly bind the
diffusing particle?  The role of such binding sites onto the protein
search for targets on DNA chain was recently investigated within a
simplified one-dimensional model \cite{Lange15}.  The GMSV allows one
to push this analysis further toward more elaborate geometric
configurations.  The related survival probability $S(t|\xo)$ satisfies
the backward diffusion equation:
\begin{subequations}
\begin{align}  \label{eq:St_diff}
\frac{\partial S(t|\xo)}{\partial t} - D \nabla^2 S(t|\xo) &= 0 \quad (\xo\in\Omega), \\
\left. \biggl(a_i S + b_i R_i \frac{\partial S}{\partial \n} \biggr) \right|_{\pa_i} & = 0 , \\  \label{eq:St_diff_c}
S(t|\xo) \bigl|_{\pa_j} & = 1   \quad (j\ne i), 
\end{align}
\end{subequations}
with the initial condition $S(t=0|\xo) = 1$.  We emphasize that this
probability characterizes the reaction events on the trap $i$; if in
turn the particle binds any absorbing sink (with $j\ne i$), it
survives forever, see Eq. (\ref{eq:St_diff_c}).  The probability
density of the reaction time is still $H(t|\xo) = - \partial
S(t|\xo)/\partial t$ but it is not normalized to $1$ given that the
reaction may never happen due to irreversible binding.

The Laplace transform reduces the diffusion equation
(\ref{eq:St_diff}) to the modified Helmholtz equation.  Rewriting this
equation for the Laplace-transformed probability density,
$\tilde{H}(p|\xo) = 1 - p \tilde{S}(p|\xo)$, one gets
\begin{subequations}  \label{eq:H_FPT_new}
\begin{align}
(p - D \nabla^2) \tilde{H}(p|\xo) &= 0 \quad (\xo\in\Omega), \\  \label{eq:H_Robin}
\left. \biggl(a_j \tilde{H} + b_j R_j \frac{\partial \tilde{H}}{\partial \n} \biggr) \right|_{\pa_j} & = a_j \delta_{ij}
\quad (j=0,\ldots,N),
\end{align}
\end{subequations}
where $a_j = 1$ and $b_j = 0$ for all $j\ne i$.  As this is a specific
case of the general boundary value problem considered in
Sec. \ref{sec:general}, its semi-analytical solution is accessible via
the GMSV.  If one is interested in finding the reaction time on a
subset $I$ of traps, the condition $a_j = 1$ and $b_j = 0$ is imposed
only for $j\notin I$, and the right-hand side of
Eq. (\ref{eq:H_Robin}) becomes $a_j 1_{j\in I}$, where $1_{j\in I}$ is
the boolean variable taking $1$ if $j\in I$ and $0$ otherwise.  When
$I = \{0,1,\ldots,N\}$, one retrieves the standard first-passage time
problem, with $a_j$ standing in the right-hand side for all $j$.  Note
also that some traps from the subset $I$ can be reflecting and thus
represent passive obstacles.  Finally, as $\tilde{H}(0|\xo)$ is the
integral of $H(t|\xo)$, it can be interpreted as the probability of
reaction, also known as the splitting probability for perfectly
reactive traps.

\subsection{Stationary diffusion of mortal particles}
\label{sec:mortal}

In the case of perfectly absorbing traps ($a_i = 1$, $b_i = 0$), the
boundary condition (\ref{eq:H_Robin}) simply reads $\tilde{H}|_{\pa_j}
= \delta_{ij}$, and the above first-passage time problem is equivalent
to stationary diffusion of ``mortal'' particles, which move from a
source on $\pa_i$ to perfect sinks on the remaining spheres $\pa_j$
and spontaneously disappear with the bulk rate $p$.  This is a very
common situation in biological and chemical diffusion-reaction
processes.  Among typical examples, one can mention: spermatozoa
moving in an aggressive medium toward an egg cell; bacteria or viruses
that can be neutralized by the immune system; cells or animals
searching for food and starving to death; proteins or RNA molecules
which can disassemble and be recycled within the cell; fluorescent
proteins diffusing toward receptors and spontaneously loosing their
signal and thus disappearing from view in single-particle tracking
experiments; excited nuclei loosing their magnetization due to
relaxation processes in nuclear magnetic resonance experiments;
diffusing radioactive nuclei that may disintegrate on their way from
the nuclear reactor core; more generally, molecules that can be
irreversibly bound to bulk constituent or be chemically transformed on
their way to catalytic sites
\cite{Yuste13,Meerson15,Grebenkov17d,Grebenkov07,Schuss19}.  For
instance, setting a constant concentration $c_0$ on the outer sphere
$\pa_0$ and zero concentration on the inner spheres $\pa_j$ describes
the diffusive flux of particles toward perfect sinks.  Alternatively,
one can impose a constant flux on the outer sphere to model particles
constantly coming onto $\pa_0$ from the exterior space.  Similarly,
any set of inner balls can play the role of a source.  In turn,
setting Neumann condition on some inner spheres switches them to inert
obstacles, whereas Robin condition describes an intermediate behavior.
The diffusive flux onto the trap $\Omega_j$ is then obtained from
Eq. (\ref{eq:dwdn}):
\begin{align}  \nonumber
J_j & = \int\limits_{\pa_j} d\s \, \left. \biggl(-D c_0 \frac{\partial \tilde{H}(p|\xo)}{\partial \n}\biggr)\right|_{\xo = \s\in \pa_j} \\
& = -\sqrt{4\pi} c_0 D R_j^2 \bigl(\F \MM^\dagger \bigr)_{00}^j   ,
\end{align}
where the matrix $\MM$ is defined by Eq. (\ref{eq:M_DtN}), and we used
the orthogonality of spherical harmonics.  Here, the components of the
vector $\F$ from Eq. (\ref{eq:F_def}) describe whether the $i$-th ball
is source or sink.  For instance, if there is a single source located
on the sphere $\pa_i$, then $f_j(\s) = \delta_{ij}$ and thus $F_{mn}^j
= \delta_{ij} \delta_{n0} \delta_{m0} \sqrt{4\pi}$ so that
\begin{equation}
J_j = - 4\pi D c_0 R_j^2 \bigl(\MM \bigr)_{00,00}^{ji}  .
\end{equation}
Expectedly, the flux is positive on traps and negative on the source.
When there is a subset of sources, then this expression is summed over
$i$ corresponding to sources.  Note that all balls can be treated as
sources, in which case particles disappear only due to the bulk rate
$p$.

As an example, let us consider two concentric spheres and assign the
outer sphere $\Omega_0$ to be a source and the inner sphere $\Omega_1$
to be a sink.  In this elementary setting, one gets an explicit
solution
\begin{align}
w(\x;q) & = c_0 \frac{i_0(q|\x|) k_0(qR_1) - k_0(q|\x|) i_0(qR_1)}{i_0(qR_0) k_0(qR_1) - k_0(qR_0) i_0(qR_1)}  \,,\\
J_1 & = \frac{4\pi c_0 D qR_0R_1}{\sinh(q(R_0-R_1))}  \,,
\end{align}
with $q = \sqrt{p/D}$.  In the limit $p\to 0$ and $R_0 \to \infty$,
one retrieves the Smoluchowski formula for the steady-state reaction
rate of a ball of radius $R_1$: $J_1 = 4\pi c_0 D R_1$.

\subsubsection*{Reaction rate}

On the other hand, the integral of $\tilde{j}(\s,p|\xo)$ from
Eq. (\ref{eq:tildej}) yields the probability flux onto the sphere
$\pa_i$ from a point source at $\xo$.  If there is a constant bulk
uptake (with concentration $c_0$), the diffusive uptake onto the trap
$\pa_i$ is given by
\begin{equation}   \label{eq:tildeJ_ip}
\overline{J}_i(p) = c_0 \int\limits_\Omega d\xo  \int\limits_{\pa_i} d\s \, \tilde{j}(\s,p|\xo)\bigr|_{\pa_i}  
 = \sqrt{4\pi} c_0 R_i^2 \overline{\J}_{00}^i ,
\end{equation}
where $\overline{\J}$ is the vector with components
$\overline{J}_{mn}^i$ given by Eq. (\ref{eq:tildeJ}) after an explicit
integration of the elements of the vector $\J$ over the starting point
$\xo$.  This is the amount of molecules (e.g., in mole) that have not
disappeared in the bulk and come to the trap $\Omega_i$.  This
quantity can also be interpreted as the Laplace transform of the
time-dependent reaction rate $J_i(t)$ for the $i$-th trap, if the
molecules were initially distributed uniformly in the domain (with
concentration $c_0$).  The Laplace-transformed total reaction rate is
then obtained by summing these diffusive fluxes:
\begin{equation}  \label{eq:Jtotal}
\tilde{J}(p) = \sum\limits_{i=0}^N \overline{J}_i(p) .
\end{equation}
For the exterior problem, the term $i=0$ corresponding to the outer
boundary is removed.  In this case, $\tilde{J}(p) \propto 1/p$ as
$p\to 0$, and the proportionality coefficient is the steady-state
reaction rate in the long-time limit.

For instance, for the exterior problem for a single sphere, one easily
gets from Eq. (\ref{eq:tildeJ}) that $\overline{\J}_{00}^i =
\sqrt{4\pi} k_1(qR_1)/(qk_0(qR_1))$, from which
\begin{equation}  \label{eq:Jtilde_Smol}
\tilde{J}_{\rm sm}(p) \equiv \overline{J}_1(p) = 4\pi c_0 D R_1 \biggl(\frac{1}{p} + \frac{R_1}{\sqrt{pD}}\biggr).
\end{equation}
This is the Laplace transform of the classical Smoluchowski rate on
the perfectly reactive sphere \cite{Smoluchowski17}:
\begin{equation}  \label{eq:J_Smol}
J_{\rm sm}(t) = 4\pi c_0 D R_1 \bigl(1 + R_1/\sqrt{\pi D t}\bigr) .
\end{equation}

\begin{figure}
\begin{center}
\includegraphics[width=88mm]{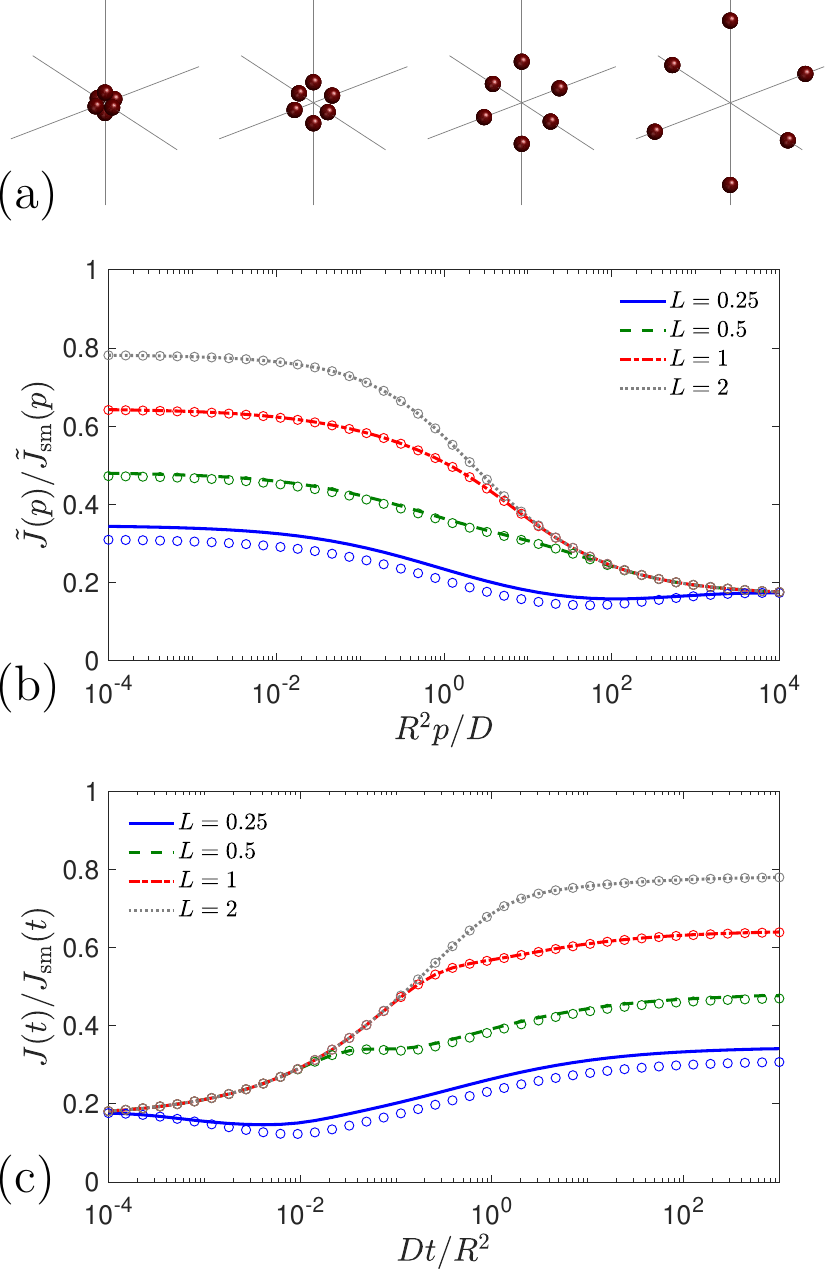}
\end{center}
\caption{
{\bf (a)} Four configurations of 6 perfect sinks of radius $\rho =
1/6$ located on the axes at distance $L$ from the origin, with $L =
0.25, 0.5, 1, 2$.  {\bf (b)} Laplace-transformed total flux
$\tilde{J}(p)$ onto 6 sinks, normalized by $\tilde{J}_{\rm sm}(p)$
from Eq. (\ref{eq:Jtilde_Smol}) for the unit sphere ($R = 1$).  Solid
lines show $\tilde{J}(p)$ computed via Eq. (\ref{eq:Jtotal}) with the
truncation order $\nmax = 2$; symbols show the results obtained with
$\nmax = 0$ (monopole approximation).  {\bf (c)} The corresponding
total fluxes $J(t)$, obtained via the numerical inversion of the
Laplace transform by the Talbot algorithm, which is normalized by
$J_{\rm sm}(t)$ from Eq. (\ref{eq:J_Smol}) for the unit sphere.  }
\label{fig:Jp_6balls}
% A_TraytakP_rates_fig3();  % (c)
% A_TraytakP_rates_fig2();  % (b)
% A_TraytakP_show_S(S);     % (a)
% export_fig('balls....eps', '-eps');
\end{figure}

We illustrate the effect of diffusion screening between traps
onto the reaction rate by considering several configurations of 6
identical perfect traps of radius $\rho = 1/6$ located along the axes
at distance $L$ from the origin (Fig. \ref{fig:Jp_6balls}(a)).
Figure \ref{fig:Jp_6balls}(b) shows the Laplace-transformed reaction
rate $\tilde{J}(p)$, which is normalized by the above Smoluchowski
rate $\tilde{J}_{\rm sm}(p)$ for a single spherical trap of radius
$R_1 = 6\rho = 1$.  In the limit of $p\to 0$ (no bulk reaction), the
curves tend to constants, indicating the common behavior $\tilde{J}(p)
\propto 1/p$.  As $L$ increases, the traps become more distant and
compete less for diffusing particles so that the reaction rate
increases.  Moreover, the particular choice $\rho = R_1/6$ ensures
that the ratio $\tilde{J}(0)/\tilde{J}_{\rm sm}(0)$ approaches $1$ as
$L\to\infty$: 6 very distant balls of radius $\rho$ trap the particles
as efficiently as a single trap of radius $6\rho$.  This is a
reminiscent feature of diffusion-limited reactions and of the
Smoluchowski rate, which is proportional to $R_1$ in the limit $p\to
0$.

In contrast, the opposite limit $p\to\infty$ corresponds to the
short-time behavior of the reaction rate.  As particles diffuse on
average over a distance $\sqrt{Dt}$, the 6 balls trap first the
particles in their close vicinity and thus do not compete.  As a
consequence, the total reaction rate does not depend on the distance
$L$ (if $L$ exceeds $\sqrt{Dt}$), as clearly seen on
Fig. \ref{fig:Jp_6balls}.  Moreover, in this limit, the second term
dominates in Eq. (\ref{eq:Jtilde_Smol}), and the reaction rate is
proportional to the squared radius that explains 6 times smaller limit
of $\tilde{J}(p)/\tilde{J}_{\rm sm}(p)$ as $p\to\infty$.

Figure \ref{fig:Jp_6balls}(c) illustrates these results in time domain
by showing the total flux $J(t)$, which is obtained via a numerical
Laplace transform inversion of $\tilde{J}(p)$ and then normalized by
$J_{\rm sm}(t)$ from Eq. (\ref{eq:J_Smol}).  At long times
(corresponding to $p\to 0$), the total flux reaches its steady-state
limit.  At short times (corresponding to $p\to \infty$), all curves
reach the same level $1/6$, which is the ratio between the total
surface area of 6 balls of radius $\rho = 1/6$ and the total surface
area of a single ball of radius $R_1 = 6\rho$.

Finally, we note that the reaction rates on Fig. \ref{fig:Jp_6balls}
were obtained by truncating matrices up to the order $\nmax = 2$.  As
we dealt with matrices of size $6(2+1)^2 \times 6(2+1)^2 = 54\times
54$, all curves were obtained within less than a second on a standard
laptop.  Remarkably, the use of the lowest truncation order $\nmax =
0$ yielded very accurate results (shown by symbols) when the traps are
well separated (i.e., $L \gg \rho$).  But even for close traps ($L =
0.25$), the error was not significant.  From our experience, this is a
common situation for exterior problems.  For interior problems, the
quality of the monopole approximation is usually lower.

\subsection{Advantages and limitations}
\label{sec:advantages}

As discussed in Sec. \ref{sec:intro}, different numerical methods have
been applied for solving boundary value problems for the modified
Helmholtz equation.  In contrast to these conventional methods, the
GMSV relies on the local spherical symmetries of perforated domains
made of non-overlapping balls.  In other words, the solution $w(\x;q)$
is decomposed on the basis functions $\psi_{mn}^{\pm}$, which are
written in local spherical coordinates and thus respect {\it locally}
the symmetry of the corresponding trap.  As a consequence, such
decompositions can often be truncated after few terms and still yield
accurate results.  An important advantage of the method is that the
dependence on $\x$ is analytical and explicit: once the coefficients
are found numerically, the solution and its spatial derivatives can be
easily calculated (and refined) at any set of points.  Moreover,
integrals of the solution over spherical boundaries or balls can be
found analytically with the help of re-expansions (see Appendix
\ref{sec:integral}).  The meshless character of the GMSV makes it
an alternative to the method of fundamental solutions (see
\cite{Lin16} and references therein).

Another important advantage of this method is the possibility of
solving {\it exterior} problems (when $\Omega_0 = \R^3$), which are
particularly difficult from the numerical point of view.  In fact, a
practical implementation of standard discretization schemes such as
finite difference or finite elements methods would require introducing
an artificial outer boundary to deal with a finite volume.  An outer
boundary is also needed in Monte Carlo simulations due to the
transient character of the three-dimensional Brownian motion.  In
contrast, the present approach does not require any outer boundary
because the solution is constructed on the appropriate basis functions
that vanish at infinity.  Exterior problems are actually simpler than
interior ones, as there is no need to impose boundary condition on the
outer boundary $\pa_0$.  In this light, the present approach is a
rather unique numerical tool to deal with various exterior boundary
value problems.

Finally, the GMSV opens access to such fundamental entities as the
Green function $G(\x,\xo;q)$, the Laplace operator $\nabla^2$, and the
Dirichlet-to-Neumann operator $\M_p$.  For instance, the eigenbasis of
the Laplace operator yields spectral decompositions of solutions of
diffusion and wave equations.  In turn, the eigenbasis of the
Dirichlet-to-Neumann operator allows one to deal with inhomogeneous
reactivity on traps \cite{Grebenkov19b}.  The spectral properties of
both operators in perforated domains will be investigated in a
separate paper.

As any numerical technique, the proposed method has its limitations
from the numerical point of view.  For the truncation order $\nmax$,
there are $(\nmax+1)^2$ basis functions $\psi_{mn}^{\pm}$ for each
ball so that the total number of unknown coefficients $A_{mn}^i$ for a
domain with $N$ traps is $N(\nmax+1)^2$ for the exterior problem and
$(N+1)(\nmax+1)^2$ for the interior problem.  Their numerical
computation involves the construction and inversion of the matrix $\W$
of size $N(\nmax+1)^2 \times N(\nmax+1)^2$.  To speed up the
construction of the matrix elements, we adapted recurrence relations
for addition theorems from Ref. \cite{Chew92}, see Appendix
\ref{sec:recurrence}.  However, the direct inversion of $\W$ becomes
very time-consuming when the number of traps $N$ and/or the truncation
order $\nmax$ grow.  As some re-expansion formulas have a limited
validity range (see Appendix \ref{sec:AW}), their truncations should
include more basis functions when the balls are close to each other.
In other words, computations for dense packings of balls need larger
$\nmax$.  In such cases, one has to resort to iterative methods (see
discussion in Ref. \cite{Grebenkov19}).  Significant numerical
improvements of this approach can be achieved by using fast multipole
methods
\cite{Gumerov02,Gumerov05,Coifman93,Darve90,Epton95,Greengard97,Cheng06,Hesford10}.
Note also that the size of the matrices is reduced to $N(\nmax+1)
\times N(\nmax+1)$ in the case of axiosymmetrical problems by using
special forms of re-expansion theorems \cite{Traytak08}.
Another drawback of the method is that the parameter $q$ enters in all
matrix elements that requires recomputing these matrices for each
value of $q$.  This is inconvenient for a numerical computation of the
inverse Laplace transform of a solution of the modified Helmholtz
equation in order to get back to time domain (see discussion in
Appendix \ref{sec:complex_q} and in Ref. \cite{Gordeliy09}).  In turn,
one can still analyze the short-time and long-time asymptotic
behaviors by considering the large-$q$ and small-$q$ limits,
respectively.

\subsection{Extensions}
\label{sec:extensions}

The GMSV can be further extended.  For instance, we assumed that $a_i$
and $b_i$ are nonnegative constants.  This assumption can be relaxed
by considering $a_i$ and $b_i$ as continuous nonnegative functions on
each sphere $\pa_i$.  The overall method is still applicable, even
though its practical implementation is more elaborate.  In fact, the
matrix elements $W_{mn,kl}^{j,i}$ and $F_{mn}^j$ will involve the
scalar products of the form $(Y_{mn}, a_i Y_{kl})_{L_2(\pa_i)}$ and
$(Y_{mn}, b_i Y_{kl})_{L_2(\pa_i)}$ that need to be computed.  Even so
such computations are rather standard (see, e.g.,
\cite{Grebenkov19b}), we do not discuss this general setting in
detail.  One can also consider other canonical domains (e.g.,
cylinders) for which re-expansion theorems are available
\cite{Erofeenko}.

Another direction for extensions consists in considering more
sophisticated kinetics on the boundary.  The Robin boundary condition
employed in the present work describes irreversible binding/reaction
on an impermeable boundary (e.g., of a solid catalyst).  In many
biological and technological applications, the boundary is a
semi-permeable membrane that separates liquid and/or gaseous phases
(e.g., intracellular and extracellular compartments).  To describe
diffusion in both phases, one can introduce two Green functions
(satisfying the modified Helmholtz equation in each phase) and couple
them via two exchange boundary conditions.  Expanding the Green
function over basis functions in each phase, one can establish the
system of linear algebraic equations on their coefficients, in a very
similar way as done in Sec. \ref{sec:general}, see
Ref. \cite{Grebenkov19} for a detailed implementation in the case of
the Laplace equation.  Yet another option is to allow for reversible
binding to the balls.  In the Laplace domain, the reversible binding
can be implemented by replacing the constant reactivity by an
effective $p$-dependent reactivity
\cite{Agmon90,Tachiya80,Agmon84,Kim99,Prustel13,Grebenkov19k}.  In
other words, the coefficients $a_i$ become $p$-dependent but the whole
method remains applicable without any change.  Note that each trap can
be characterized by its own dissociation rate.  This extension allows
one to investigate the role of immobile buffering molecules in
signalling processes, DNA search processes, and gene regulations, as
well as many other chemical reactions (see
\cite{Li09,Benichou09,Bressloff13,Lange15} and references therein).

\section{Conclusion}
\label{sec:conclusion}

The GMSV was broadly employed for solving boundary value problems for
the Laplace and ordinary Helmholtz equations in different disciplines
ranging from electrostatics to hydrodynamics and scattering theory.
Quite surprisingly, applications of this powerful method to the
modified Helmholtz equation, which plays the crucial role for
describing diffusion-reaction processes in chemical physics, are much
less developed.  In the present paper, we described a general
analytical and numerical framework for solving such problems in
perforated domains made of non-overlapping balls.  In particular, we
provided a semi-analytical solution $w(\x;q)$, in which the dependence
on the point $\x$ enters {\it analytically} through explicitly known
basis functions $\psi_{mn}^{\pm}$, while their coefficients are
obtained {\it numerically} by truncating and solving the established
system of linear algebraic equations.  The high numerical efficiency
of this approach relies on exploiting the local symmetries of the
spherical traps and using the most natural basis functions.

We applied this method to derive a semi-analytical representation of
the Green function that determines various characteristics of
non-stationary diffusion among partially reactive traps such as the
probability flux density, the reaction rate, the survival probability,
and the associated probability density of the reaction time.  We also
showed how this method can be adapted to obtain the eigenvalues and
eigenfunctions of the Laplace operator and of the Dirichlet-to-Neumann
operator.  These operators play an important role in mathematical
physics and have been applied in a variety of disciplines, including
chemical physics.

We described several applications of this technique such as the
first-passage properties and stationary diffusion of mortal particles.
In particular, we checked the conjecture that reflecting obstacles
cannot speed up the exit from the center of a ball.  Interestingly,
the presence of even large obstacles had a minor effect on the
distribution of the first-exit time.  We also discussed how the mutual
distance between absorbing traps affects the reaction rate.  This
discussion brings complementary insights onto the role of diffusion
screening (or interaction) onto the reaction rate, which was
thoroughly investigated in the steady-state limit ($t \to \infty$) but
remains less known in the time-dependent regime.  More generally, the
developed framework provides a solid theoretical ground and efficient
numerical tool for studying diffusion-controlled reactions in various
media that can be modeled by spherical traps and obstacles.

\begin{acknowledgments}
The author thanks Prof. S. D. Traytak for fruitful discussions.
\end{acknowledgments}

\section*{Data Availability Statement}

The data that support the findings of this study are available from
the corresponding author upon reasonable request.

\appendix

\section{Technical derivations}

\subsection{Re-expansion theorems}
\label{sec:re-expansion}

The efficiency of the GMSV relies on re-expansion (or addition)
theorems that allow one to represent the basis functions
$\psi_{mn}^{\pm}(qr_j,\theta_j,\phi_j)$, written in the local
spherical coordinates $(r_j,\theta_j,\phi_j)$ associated with the ball
$\Omega_j$, in terms of the basis functions
$\psi_{mn}^{\pm}(qr_i,\theta_i,\phi_i)$ in the local spherical
coordinates $(r_i,\theta_i,\phi_i)$ associated with the ball
$\Omega_i$.  We first recall three re-expansion theorems for the
ordinary Helmholtz equation and then adapt them to the modified
Helmholtz equation.

\begin{figure}
\begin{center}
\includegraphics[width=88mm]{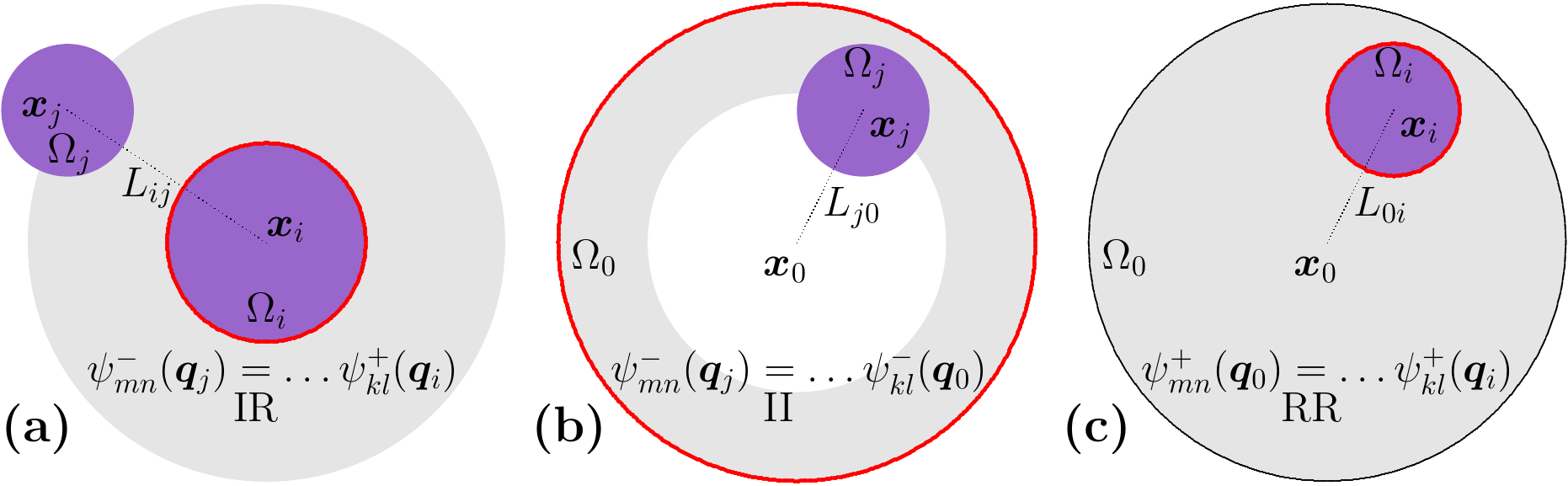}
\end{center}
\caption{
Three translational re-expansion theorems that we use to decompose any
basis function $\psi_{mn}^{\pm}(\qq_j) =
\psi_{mn}^{\pm}(qr_j,\theta_j,\phi_j)$ in the local spherical
coordinates $(r_j,\theta_j,\phi_j)$ centered at $\x_j$, on the basis
functions $\psi_{kl}^{\pm}(qr_i,\theta_i,\phi_i)$ in the local
spherical coordinates $(r_i,\theta_i,\phi_i)$ centered at $\x_i$.
These decompositions help us to impose the boundary condition on
$\partial \Omega_i$ (shown by red circle).
{\bf (a)} The basis function $\psi_{mn}^{-}(\qq_j)$ outside the ball
$\Omega_j$ can be decomposed on regular functions
$\psi_{kl}^{+}(\qq_i)$ via the IR theorem (\ref{eq:Hadd_SR}), which is
valid in the shadowed (gray) region, in particular, on $\pa_i$.
{\bf (b)} In turn, to impose the boundary condition on the outer
boundary $\pa_0$, the basis function $\psi_{mn}^{-}(\qq_j)$ outside
the ball $\Omega_j$ is decomposed on irregular functions
$\psi_{kl}^{-}(\qq_0)$ via the II theorem (\ref{eq:Hadd_SS}), which is
valid in the shadowed region, in particular, on $\pa_0$.
{\bf (c)} The basis function $\psi_{mn}^{+}(\qq_0)$ inside the ball
$\Omega_0$ is decomposed on regular functions $\psi_{kl}^{+}(\qq_i)$
via the RR theorem (\ref{eq:Hadd_RR}), which is valid anywhere, in
particular, on $\pa_i$.}
\label{fig:TAT_types}
\end{figure}

Let us denote by $\L_{ij} = \x_i - \x_j$ the vector connecting the
centers of balls $j$ and $i$, and $(L_{ij}, \Theta_{ij},
\Phi_{ij})$ are the spherical coordinates of the vector $\L_{ij}$
(Fig. \ref{fig:TAT_types}):
\begin{equation} 
\begin{split}
x_i & = x_j + L_{ij}\sin \Theta_{ij} \cos \Phi_{ij},  \\
y_i & = y_j + L_{ij}\sin \Theta_{ij} \sin \Phi_{ij},  \\
z_i & = z_j + L_{ij}\cos \Theta_{ij} . \\
\end{split}
\end{equation}
For the basis functions $\tilde{\psi}_{mn}^{\pm}$ of the ordinary
Helmholtz equation (\ref{eq:Helm_ordinary}), three translational
re-expansion theorems are \cite{Hobson,Friedman54,Epton95,Erofeenko}:

(i) regular-regular (RR) addition theorem:
\begin{equation}
\label{eq:Hadd_RR_ord}
\tilde{\psi}_{mn}^{+}(q r_j,\theta_j,\phi_j) = \sum\limits_{k,l} \tilde{U}^{(+j,+i)}_{mn,kl} \, \tilde{\psi}_{kl}^{+}(q r_i,\theta_i,\phi_i) ,
\end{equation}
where the matrix elements of the translation operator are
\begin{equation}  \label{eq:HU_RR_ord}
\tilde{U}^{(+j,+i)}_{mn,kl} = \sum\limits_{\nu=|l-n|}^{l+n} i^{\nu+l-n} \, b^\nu_{nmlk} \, \tilde{\psi}_{(m-k)\nu}^{+}(q L_{ij}, \Theta_{ij}, \Phi_{ij}) ,
\end{equation}
in which
\begin{align}  
b_{nmlk}^\nu & = (-1)^{k} \sqrt{4\pi (2l+1)(2n+1)/(2\nu+1)} \\  \nonumber 
& \times \langle n, l, 0, 0 | n,l,\nu, 0 \rangle ~ \langle n,l,m, -k|n,l,\nu,m-k\rangle , 
\end{align}
with $\langle j_1,j_2,m_1,m_2|j_1,j_2,j,m\rangle$ being the
Clebsch-Gordan coefficients (see Ref. \cite{Abramowitz}, Section
27.9);

(ii) irregular-regular (IR) addition theorem:
\begin{equation}  \label{eq:Hadd_SR_ord}
\tilde{\psi}_{mn}^{-}(q r_j,\theta_j,\phi_j) = \sum\limits_{k,l} \tilde{U}^{(-j,+i)}_{mn,kl} \,  \tilde{\psi}_{kl}^{+}(q r_i,\theta_i,\phi_i) 
\end{equation}
(for $r_i < L_{ij}$), where
\begin{equation}  \label{eq:HU_SR_ord}
\tilde{U}^{(-j,+i)}_{mn,kl} = \sum\limits_{\nu=|l-n|}^{l+n} i^{\nu+l-n} b^\nu_{nmlk} \, \tilde{\psi}_{(m-k)\nu}^{-}(q L_{ij}, \Theta_{ij},\Phi_{ij}) ;
\end{equation}

(iii) irregular-irregular (II) addition theorem:
\begin{equation}  \label{eq:Hadd_SS_ord}
\tilde{\psi}_{mn}^{-}(q r_j,\theta_j,\phi_j) = \sum\limits_{l,k} \tilde{U}^{(-j,-i)}_{mn,kl} \, \tilde{\psi}_{kl}^{-}(q r_i,\theta_i,\phi_i) 
\end{equation}
(for $r_i > L_{ij}$), where
\begin{equation}  \label{eq:HU_SS_ord}
\tilde{U}^{(-j,-i)}_{mn,kl} = \hspace*{-1.5mm} \sum\limits_{\nu=|l-n|}^{l+n} \hspace*{-1.5mm} 
i^{\nu+l-n}  b^l_{nm\nu(m-k)}  \tilde{\psi}_{(m-k)\nu}^{+}(q L_{ij}, \Theta_{ij},\Phi_{ij}) .
\end{equation}
We recall that the basis functions for the ordinary Helmholtz equation
are
\begin{subequations}
\begin{align}  \label{eq:psi_plus_tilde}
\tilde{\psi}_{mn}^{+}(qr_i,\theta_i,\phi_i) & = j_n(qr_i) \, Y_{mn}(\theta_i,\phi_i) , \\
\tilde{\psi}_{mn}^{-}(qr_i,\theta_i,\phi_i) & = h_n^{(1)}(qr_i) \, Y_{mn}(\theta_i,\phi_i), 
\end{align}
\end{subequations}
where $j_n(z) = \sqrt{\pi/(2z)}\, J_{n+1/2}(z)$ and $h_n^{(1)}(z) =
\sqrt{\pi/(2z)} \, H_{n+1/2}^{(1)}(z)$ are the spherical Bessel and Hankel
functions of the first kind (note that $h_n^{(1)}(z) = j_n(z) + i y_n(z)$).

For convenience, the first sign in the superscript of the matrix
elements $\tilde{U}_{mn,kl}^{(\pm j,\pm i)}$ denotes the type of the
basis function $\tilde{\psi}_{mn}^{\pm}(qr_j,\theta_j,\phi_j)$ to be
expanded ($+$ for regular and $-$ for irregular one), whereas the
second sign refers to the type of the basis functions
$\tilde{\psi}_{kl}^{\pm}(qr_i,\theta_i,\phi_i)$ over which the
expansion is provided.  While the first addition theorem holds for any
values of $r_i$ and $r_j$, the second and the third ones are
applicable for $r_i < L_{ij}$ and $r_i > L_{ij}$, respectively (see
Fig. \ref{fig:TAT_types}).  In the above expressions, we used tilde to
outline that the involved basis functions and the matrix elements
correspond to the ordinary Helmholtz equation.

Replacing $q$ by $iq$ and using the relations
\begin{equation}
\label{eq:jn_in}
j_n(iz) = i^n i_n(z),  \qquad h_n^{(1)}(iz) = - i^{-n} k_n(z) ,
\end{equation}
one gets three translational re-expansion (or addition) theorems for
the modified Helmholtz equation (Fig. \ref{fig:TAT_types}):

(i) regular-regular (RR) addition theorem:
\begin{equation}
\label{eq:Hadd_RR}
\psi_{mn}^{+}(q r_j,\theta_j,\phi_j) = \sum\limits_{k,l} U^{(+j,+i)}_{mn,kl} \, \psi_{kl}^{+}(q r_i,\theta_i,\phi_i) ,
\end{equation}
with 
\begin{equation}  \label{eq:HU_RR}
U^{(+j,+i)}_{mn,kl} = \hspace*{-2mm}\sum\limits_{\nu=|l-n|}^{l+n} \hspace*{-1.5mm} 
(-1)^{\nu+n+l} \, b^\nu_{nmlk} \, \psi_{(m-k)\nu}^{+}(q L_{ij}, \Theta_{ij}, \Phi_{ij}) .
\end{equation}

(ii) irregular-regular (IR) addition theorem:
\begin{equation}  \label{eq:Hadd_SR}
\psi_{mn}^{-}(q r_j,\theta_j,\phi_j) = \sum\limits_{k,l} U^{(-j,+i)}_{mn,kl} \,  \psi_{kl}^{+}(q r_i,\theta_i,\phi_i) 
\end{equation}
(for $r_i < L_{ij}$), where
\begin{equation}  \label{eq:HU_SR}
U^{(-j,+i)}_{mn,kl} = \sum\limits_{\nu=|l-n|}^{l+n} (-1)^{l} b^\nu_{nmlk} \, \psi_{(m-k)\nu}^{-}(q L_{ij}, \Theta_{ij},\Phi_{ij}) .
\end{equation}

(iii) irregular-irregular (II) addition theorem:
\begin{equation}  \label{eq:Hadd_SS}
\psi_{mn}^{-}(q r_j,\theta_j,\phi_j) = \sum\limits_{l,k} U^{(-j,-i)}_{mn,kl} \, \psi_{kl}^{-}(q r_i,\theta_i,\phi_i) 
\end{equation}
(for $r_i > L_{ij}$), where
\begin{equation}  \label{eq:HU_SS}
U^{(-j,-i)}_{mn,kl} = \hspace*{-1.5mm} \sum\limits_{\nu=|l-n|}^{l+n}  \hspace*{-1mm} (-1)^\nu
 b^l_{nm\nu(m-k)} \, \psi_{(m-k)\nu}^{+}(q L_{ij}, \Theta_{ij},\Phi_{ij}) .
\end{equation}
Importantly, all the matrix elements can be found from recurrence
relations that considerably speed up their computation (see
Appendix \ref{sec:recurrence}).

\subsection{Matrix elements $W_{mn,kl}^{j,i}$}
\label{sec:AW}

In this Appendix, we derive the explicit formulas for the matrix
elements $W_{mn,kl}^{j,i}$ defined in Eq. (\ref{eq:W_def}).  Even
though one could in principle compute these elements by numerical
integration, the derived exact explicit expressions significantly
improve the speed and accuracy of the semi-analytical solution.

The matrix elements are particularly simple for $i = j$ because the
basis solution $\psi_{mn}^{\pm}(qr_i,\theta_i,\phi_i)$ is written in
the local spherical coordinates associated to the boundary $\pa_i$.
For $i = 1,2,\ldots,N$, one has
\begin{align*}
& \left.\left(a_i + b_i R_i \frac{\partial }{\partial \n}\right) \psi_{mn}^{-}(qr_i,\theta_i,\phi_i)\right|_{\pa_i} \\
& = \biggl(a_i k_n(q R_i) - b_i R_i \, q k'_n(q R_i) \biggr) Y_{mn}(\theta_i,\phi_i),
\end{align*}
where prime denotes the derivative with respect to the argument, and
the sign minus appeared from the orientation of the normal derivative:
$\partial/\partial \n = - \partial/\partial r_i$.  The scalar product
with $Y_{kl}$ yields
\begin{equation}
W_{mn,kl}^{i,i} = \delta_{nl} \delta_{mk} \bigl(a_i k_n(q R_i) - b_i R_i \, q k'_n(q R_i) \bigr),
\end{equation}
due to the orthonormality of the spherical harmonics.  Similarly, one
gets for the outer boundary ($i = j = 0$):
\begin{equation}
W_{mn,kl}^{0,0} = \delta_{nl} \delta_{mk} \bigl(a_0 i_n(q R_0) + b_0 R_0 \, q i'_n(q R_0) \bigr).
\end{equation}

The major difficulty consists in computing the matrix elements
$W_{mn,kl}^{j,i}$ for $i\ne j$ as one needs to re-expand the basis
functions $\psi_{mn}^{\pm}(qr_j,\theta_j,\phi_j)$ in terms of the
basis functions $\psi_{mn}^{\pm}(qr_i,\theta_i,\phi_i)$ with the aid
of the re-expansion theorems (Sec. \ref{sec:re-expansion}).  We start
with the case $i,j = 1,2,\ldots,N$, for which the irregular-regular
addition theorem is applied:
\begin{align*}
& \left.\left(a_i + b_i R_i \frac{\partial }{\partial \n}\right) \psi_{mn}^{-}(qr_j,\theta_j,\phi_j)\right|_{\pa_i} \\
& = \sum\limits_{k,l} U_{mn,kl}^{(-j,+i)} 
\left.\left(a_i + b_i R_i \frac{\partial }{\partial \n}\right) \psi_{kl}^{+}(qr_i,\theta_i,\phi_i)\right|_{\pa_i}  \\
& = \sum\limits_{k,l} U_{mn,kl}^{(-j,+i)} \biggl(a_i i_l(q R_i) - b_i R_i \, q i'_l(q R_i) \biggr) Y_{kl}(\theta_i,\phi_i).
\end{align*}
The scalar product of this expression with $Y_{kl}$ yields
\begin{equation}
W_{mn,kl}^{j,i} = U_{mn,kl}^{(-j,+i)} \bigl(a_i i_{l}(q R_i) - b_i R_i \, q i'_{l}(q R_i) \bigr).
\end{equation}

When $i = 0$ and $j = 1,2,\ldots, N$, one uses the irregular-irregular
addition theorem:
\begin{align*}
& \left.\left(a_0 + b_0 R_0 \frac{\partial }{\partial \n}\right) \psi_{mn}^{-}(qr_j,\theta_j,\phi_j)\right|_{\pa_0} \\
& = \sum\limits_{k,l} U_{mn,kl}^{(-j,-0)} 
\left.\left(a_0 + b_0 R_0 \frac{\partial }{\partial \n}\right) \psi_{kl}^{-}(qr_0,\theta_0,\phi_0)\right|_{\pa_0}  \\
& = \sum\limits_{k,l} U_{mn,kl}^{(-j,-0)} \biggl(a_0 k_l(q R_0) + b_0 R_0 \, q k'_l(q R_0) \biggr) Y_{kl}(\theta_0,\phi_0),
\end{align*}
from which
\begin{equation}
W_{mn,kl}^{j,0} = U_{mn,kl}^{(-j,-0)} \bigl(a_0 k_{l}(q R_0) + b_0 R_0 \, q k'_{l}(q R_0) \bigr).
\end{equation}

Finally, when $i = 1,2,\ldots,N$ and $j = 0$, one uses the
regular-regular addition theorem:
\begin{align*}
& \left.\left(a_i + b_i R_i \frac{\partial }{\partial \n}\right) \psi_{mn}^{+}(qr_0,\theta_0,\phi_0)\right|_{\pa_i} \\
& = \sum\limits_{k,l} U_{mn,kl}^{(+0,+i)} 
\left.\left(a_i + b_i R_i \frac{\partial }{\partial \n}\right) \psi_{kl}^{+}(qr_i,\theta_i,\phi_i)\right|_{\pa_i}  \\
& = \sum\limits_{k,l} U_{mn,kl}^{(+0,+i)} \biggl(a_i i_l(q R_i) - b_i R_i \, q i'_l(q R_i) \biggr) Y_{kl}(\theta_i,\phi_i),
\end{align*}
from which
\begin{equation}
W_{mn,kl}^{0,i} = U_{mn,kl}^{(+0,+i)} \bigl(a_i i_{l}(q R_i) - b_i R_i \, q i'_{l}(q R_i) \bigr).
\end{equation}

In summary, the matrix $\W$ is formed by $(N+1) \times (N+1)$ blocks
corresponding to indices $i,j = 0,1,\ldots,N$.  It is convenient to
represent this matrix as
\begin{equation}   \label{eq:HAW}
\W = \q + \U \p ,
\end{equation}
with
\begin{equation}  \label{eq:U_def} \small
\U = \left(\begin{array}{c c c c c} \O & \U^{(+0,+1)} & \U^{(+0,+2)} & \cdots & \U^{(+0,+N)} \\  
\U^{(-1,-0)} & \O & \U^{(-1,+2)} & \cdots & \U^{(-1,+N)} \\  
\U^{(-2,-0)} & \U^{(-2,+1)} & \O & \cdots & \U^{(-2,+N)} \\  
\cdots  & \cdots  & \cdots& \cdots& \cdots \\
\U^{(-N,-0)} & \U^{(-N,+1)} & \U^{(-N,+2)} & \cdots & \O \\   \end{array} \right), 
\end{equation}
where the block matrices $\U^{(\pm j,\pm i)}$ are formed by the
elements $U_{mn,kl}^{(\pm j,\pm i)}$ given above, and $\p$ and $\q$
are block-diagonal matrices with the elements
\begin{subequations}  \label{eq:pq_def}
\begin{align}  \label{eq:p_def}
\bigl(\p\bigr)_{mn,kl}^{ij} & = a_i \bigl(\tilde{\p}\bigr)_{mn,kl}^{ij} + b_i R_i \bigl(\tilde{\p}'\bigr)_{mn,kl}^{ij} \,, \\  \label{eq:q_def}
\bigl(\q\bigr)_{mn,kl}^{ij} & = a_i \bigl(\tilde{\q}\bigr)_{mn,kl}^{ij} + b_i R_i \bigl(\tilde{\q}'\bigr)_{mn,kl}^{ij} \,,
\end{align}
\end{subequations}
with
\begin{subequations}  \label{eq:pqtilde}
\begin{align}  
\bigl(\tilde{\p}\bigr)_{mn,kl}^{ij} & = \delta_{ij} \delta_{nl} \delta_{mk} \begin{cases} k_n(qR_0)  \quad (i = 0), \cr
i_n(qR_i)   \hskip 4.5mm (i > 0), \end{cases}   \\   \label{eq:ptilde}
\bigl(\tilde{\p}'\bigr)_{mn,kl}^{ij} & = \delta_{ij} \delta_{nl} \delta_{mk}  \begin{cases} q k'_n(qR_0)  \hskip 5mm (i = 0), \cr
- q i'_n(qR_i)   \quad (i > 0), \end{cases} \\
\bigl(\tilde{\q}\bigr)_{mn,kl}^{ij} & = \delta_{ij} \delta_{nl} \delta_{mk} \begin{cases} i_n(qR_0)  \quad (i = 0), \cr
k_n(qR_i)   \hskip 3.5mm (i > 0), \end{cases}   \\    \label{eq:qtilde}
\bigl(\tilde{\q}'\bigr)_{mn,kl}^{ij} & = \delta_{ij} \delta_{nl} \delta_{mk} \begin{cases} q i'_n(qR_0)  \hskip 6.5mm (i = 0), \cr
- q k'_n(qR_i)   \quad (i > 0). \end{cases}  
\end{align}
\end{subequations}
The first block row and the first block column of the matrix $\U$ are
different from the other blocks as they are related to the outer
boundary $\pa_0$.  For the exterior problem, all $A_{mn}^0 \equiv 0$
and thus the matrices $\U$ and $\W$ are reduced by removing the first
block row and block column.

We can thus combine the above relations in a single expression:
\begin{align} \nonumber
\biggl(a_i + b_i R_i \frac{\partial}{\partial \n}\biggr) & \psi_{mn}^{\epsilon_j}(qr_j,\theta_j,\phi_j) \biggr|_{\pa_i}   \\   \label{eq:ab_psi}
& = \sum\limits_{k,l} (\q + \U \p)_{mn,kl}^{ji} \, Y_{kl}(\theta_i,\phi_i) .
\end{align}
In particular, the restrictions of
$\psi_{mn}^{\epsilon_j}(qr_j,\theta_j,\phi_j)$ and of its normal
derivative onto $\pa_i$ involve the matrices $\tilde{\p}, \tilde{\q}$
and $\tilde{\p}', \tilde{\q}'$, respectively, see Eqs. (\ref{eq:pq_def}).

As the vectors $\L_{ij}$ and $\L_{ji}$ have opposite directions, one
has
\begin{equation}
\psi_{mn}^\pm(qL_{ij},\Theta_{ij},\Phi_{ij}) = (-1)^n \psi_{mn}^\pm(qL_{ji},\Theta_{ji},\Phi_{ji}) .
\end{equation}
Using the symmetry properties of Clebsch-Gordan coefficients (see
Ref. \cite{Abramowitz}, Section 27.9), one can show that the matrix
$\U$ is Hermitian: $\U^\dagger = \U^*$.  As a consequence, one has
\begin{equation}
\W^{\dagger,*} = \q + \p \U .
\end{equation}

\subsection{Matrix elements $F_{mn}^i$ for the Green function}
\label{sec:AF}

In order to compute the matrix elements $F_{mn}^i$ needed for the
evaluation of the Green function, we use the following expansion of
the fundamental solution (derived from Ref. \cite{Friedman54}):
\begin{widetext}
\begin{equation} \label{eq:Gf}   
\begin{split}
G_{\rm f}(\x,\xo;q) & = \begin{cases} 
\displaystyle  q \sum\limits_{n=0}^\infty \sum\limits_{m=-n}^n (-1)^m 
\psi_{(-m)n}^{+}(qr_0 ,\theta_0,\phi_0) \, \psi_{mn}^{-}(qr, \theta, \phi)  \qquad (r_0 < r),  \cr
\displaystyle  q \sum\limits_{n=0}^\infty \sum\limits_{m=-n}^n (-1)^m 
\psi_{mn}^{-}(qr_0 ,\theta_0,\phi_0) \, \psi_{(-m)n}^{+}(qr, \theta, \phi)  \qquad (r_0 > r),  \end{cases} \\
\end{split}
\end{equation}
\end{widetext}
where $(r,\theta,\phi)$ and $(r_0,\theta_0,\phi_0)$ are respectively
the spherical coordinates of $\x$ and $\xo$ with respect to the
origin.  Since the fundamental solution is translationally invariant,
i.e., $G_{\rm f}(\x,\xo;q) = G_{\rm f}(\x - \x_i,\xo - \x_i;q)$ for
any vector $\x_i$, one can apply Eqs. (\ref{eq:Gf}) to represent
$G_{\rm f}(\x,\xo;q)$ in the local spherical coordinates of the ball
$\Omega_i$ (with $i = 1,2,\ldots,N$):
\begin{equation}  \label{eq:Gf_V}
G_{\rm f}(\x,\xo;q) = q \sum\limits_{m,n} \PPsi_{mn}^{i,*}(\xo)  \, \psi_{mn}^{+}(qr_i, \theta_i, \phi_i)  \quad (r_i < L_i),
\end{equation}
in which we introduced the (row) vector $\PPsi(\xo)$ with the
components
\begin{equation} \label{eq:HV}
\PPsi_{mn}^i(\xo) = \psi_{mn}^{\epsilon_i}(q L_i, \Theta_i, \Phi_i),
\end{equation}
where $(L_i, \Theta_i, \Phi_i)$ are the local spherical coordinates of
$\xo$ with respect to the ball $\Omega_i$ for $i = 0,1,\ldots,N$
(i.e., the spherical coordinates of the vector $\xo - \x_i$).  In
addition, to replace $(-1)^m \psi_{(-m)n}$ by $\psi_{mn}^*$, we used
the following identity for the normalized spherical harmonics:
\begin{equation}  \label{eq:Y_m}
Y_{(-m)n}(\theta,\phi) = (-1)^m \, Y_{mn}^*(\theta,\phi),
\end{equation}
which follows from the identity for associated Legendre polynomials:
\begin{equation}
P_n^{-m}(x) = (-1)^m \frac{(n-m)!}{(n+m)!}\, P_n^m(x).
\end{equation}

From Eq. (\ref{eq:Gf_V}), we determine $f_i$ according to
Eq. (\ref{eq:fi_def}):
\begin{equation}
f_i = q \sum\limits_{m,n} \PPsi_{mn}^{i,*}(\xo)  \bigl(a_i i_n(qR_i) - b_i R_i \, q \, i'_n(qR_i) \bigr) Y_{mn}(\theta_i, \phi_i)  
\end{equation}
and thus
\begin{equation}
F_{mn}^i = q \PPsi_{mn}^{i,*}(\xo)  \bigl(a_i i_n(qR_i) - b_i R_i \, q \, i'_n(qR_i) \bigr).
\end{equation}
Similarly, the representation of $G_{\rm f}(\x,\xo;q)$ in the local
spherical coordinates of the ball $\Omega_0$ reads
\begin{equation}  \label{eq:Gf_V0}
G_{\rm f}(\x,\xo;q) =  q \sum\limits_{m,n} \PPsi_{mn}^{0,*}(\xo) \psi_{mn}^{-}(qr_0, \theta_0, \phi_0)  \quad (r_0 > L_0) ,
\end{equation}
from which we get
\begin{align} \nonumber
f_0 & = q \sum\limits_{m,n} \PPsi_{mn}^{0,*}(\xo) \bigl(a_0 k_n(qR_0) + b_0 R_0 \, q \, k'_n(qR_0) \bigr) \\
& \times Y_{mn}(\theta_0, \phi_0) ,
\end{align}
and thus
\begin{equation}
F_{mn}^0 = q \PPsi_{mn}^{0,*}(\xo) \bigl(a_0 k_n(qR_0) + b_0 R_0 \, q \, k'_n(qR_0) \bigr).
\end{equation}
Using the matrix $\p$ from Eq. (\ref{eq:p_def}), one can represent the
vector $\F$ as
\begin{equation}  \label{eq:F_V}
\F = q \PPsi^*(\xo) \p .
\end{equation}

\subsection{Normal derivative of the solution}
\label{sec:dwdn}

Once the coefficients $A_{mn}^i$ are found, one can easily evaluate
the solution $w(\x;q)$ and its derivatives in any point $\x$.  For
many applications, one needs to compute the restriction of the
solution and the flux onto the boundary $\pa$ which requires finding
the normal derivative of $w(\x;q)$.

Using Eq. (\ref{eq:ab_psi}) with $a_i = 1$ and $b_i R_i = 0$, one gets
immediately
\begin{equation} \label{eq:wA}
\left. w \right|_{\pa_i} =  
\sum\limits_{m,n} \bigl(\A \tilde{\W}\bigr)_{mn}^i \, Y_{mn}(\theta_i,\phi_i) ,   
\end{equation}
where
\begin{equation} \label{eq:Wtilde0}
\tilde{\W} = \tilde{\q} + \U \tilde{\p} ,
\end{equation}
with matrices $\tilde{\p}$ and $\tilde{\q}$ given by
Eqs. (\ref{eq:pqtilde}).  In turn, setting $a_i = 0$ and $b_i R_i = 1$
in Eq. (\ref{eq:ab_psi}), one has
\begin{equation} \label{eq:dwdnA}
\left. \biggl(\frac{\partial w}{\partial \n}\biggr)\right|_{\pa_i} =  
\sum\limits_{m,n} \bigl(\A \tilde{\W}'\bigr)_{mn}^i \, Y_{mn}(\theta_i,\phi_i) ,   
\end{equation}
where
\begin{equation} \label{eq:Wtilde}
\tilde{\W}' = \tilde{\q}' + \U \tilde{\p}' .
\end{equation}

As a particular application, we evaluate the normal derivative of the
Green function $G(\x,\xo;q)$.  Using the expansions (\ref{eq:gi},
\ref{eq:Gf_V}), we get for $i > 0$:
\begin{align}  \nonumber
\frac{\partial G(\x,\xo; q)}{\partial \n} \biggr|_{\pa_i} \hspace*{-1.5mm} & = - 
\sum\limits_{m,n} \biggl\{ \bigl(q \PPsi_{mn}^{i,*}(\xo) - (\A\U)_{mn}^i\bigr)  q i'_n(qR_i) \\
& - A_{mn}^i q k'_n(qR_i) \biggr\} Y_{mn}(\theta_i,\phi_i),
\end{align}
where $\PPsi_{mn}^i(\xo)$ are given explicitly by Eq. (\ref{eq:HV}),
and
\begin{align}  \nonumber
\frac{\partial G(\x,\xo; q)}{\partial \n} \biggr|_{\pa_0} & = 
\sum\limits_{m,n} \biggl\{ \bigl(q\PPsi_{mn}^{0,*}(\xo) - (\A\U)_{mn}^0\bigr)  q k'_n(qR_0) \\
& - A_{mn}^0 q i'_n(qR_0) \biggr\} Y_{mn}(\theta_0,\phi_0).
\end{align}
Using the matrices $\tilde{\p}'$ and $\tilde{\q}'$ from
Eqs. (\ref{eq:pqtilde}) and the expressions (\ref{eq:F_AW},
\ref{eq:F_V}), the above relations can be written together (for any
$i=0,1,\ldots,N$) as
\begin{equation}  \label{eq:Atildej}
\frac{\partial G(\x,\xo; q)}{\partial \n} \biggr|_{\pa_i} = - \sum\limits_{m,n} \J_{mn}^i(\xo) \, Y_{mn}(\theta_i,\phi_i)  ,
\end{equation}
where
\begin{equation} \label{eq:Jmatrix}
\J(\xo) = q \PPsi^*(\xo) \biggl(\p (\q + \U\p)^{-1} (\tilde{\q}' + \U \tilde{\p}') - \tilde{\p}'\biggr) .
\end{equation}

It is also convenient to represent the matrix in the large parentheses
as
\begin{equation} \label{eq:Midentity}
\p (\q + \U\p)^{-1}(\tilde{\q}' + \U \tilde{\p}') - \tilde{\p}' = (\q + \p \U)^{-1} (\p \tilde{\q}' - \q \tilde{\p}') .
\end{equation}
To proof this identity, both sides can be multiplied by $(\q + \p \U)$
on the left to get
\begin{equation*}
(\q + \p \U) \p (\q + \U\p)^{-1}(\tilde{\q}' + \U \tilde{\p}') - (\q + \p \U) \tilde{\p}' = \p \tilde{\q}' - \q \tilde{\p}' .
\end{equation*}
As the matrices $\p$ and $\q$ are diagonal, they commute, and one has
\begin{equation*}
(\q + \p \U) \p (\q + \U\p)^{-1} = \p (\p^{-1}\q + \U) (\q \p^{-1} + \U)^{-1} = \p,
\end{equation*}
from which the identity (\ref{eq:Midentity}) follows.  Moreover, the
last matrix in Eq. (\ref{eq:Midentity}) has a particularly simple
form:
\begin{equation}
\bigl(\p \tilde{\q}' - \q \tilde{\p}' \bigr)_{mn,kl}^{ij} = \delta_{ij} \delta_{nl} \delta_{mk} 
\frac{a_j}{q R_j^2} \,,
\end{equation}
which is easily obtained by using the Wronskian (\ref{eq:Wronskian}).
We conclude that
\begin{equation} \label{eq:J00i}
\J_{mn}^i(\xo) = \frac{a_i}{R_i^2} \biggl(\PPsi^*(\xo) (\q + \p \U)^{-1} \biggr)_{mn}^i .
\end{equation}

\subsection{Integration of the solution}
\label{sec:integral}

The re-expansion theorems allow one to easily integrate the solution
$w(\x;q)$ of the modified Helmholtz equation over balls or spheres.
In fact, one can re-expand basis functions $\psi_{mn}^{\pm}$ in
Eq. (\ref{eq:gi}) on the appropriate basis functions in the local
spherical coordinate of a ball or a sphere, over which the integral
needs to be evaluated.  After that, the integral can be evaluated
explicitly.  To illustrate this computation, we find the integral of
the solution $w(\x;q)$ over the whole domain $\Omega$, which is a more
complicated setting.  For this purpose, one needs to compute the
integrals:
\begin{equation}
\overline{\psi}_{mn}^j = \int\limits_{\Omega} d\x \, \psi_{mn}^{\epsilon_j}(qr_j, \theta_j,\phi_j)
\end{equation}
for $j=0,1,2,\ldots,N$ (we recall that $\epsilon_j = -$ for $j>0$ and
$\epsilon_0 = +$).  The following computation relies on the additivity
of the integral over $\Omega = \Omega_0 \backslash \cup_{i=1}^N
\overline{\Omega}_i$, with non-overlapping balls.  For $j > 0$, one can
split the integral over $\Omega$ into three parts:
\begin{align*}
\overline{\psi}_{mn}^j  & = \hspace*{-1mm}\int\limits_{\R^3\backslash \Omega_j} \hspace*{-1mm} d\x \, \psi_{mn}^{-}(qr_j, \theta_j,\phi_j) 
 - \hspace*{-1mm} \int\limits_{\R^3 \backslash \Omega_0} \hspace*{-1mm} d\x \, \psi_{mn}^{-}(qr_j, \theta_j,\phi_j)\\
& - \sum\limits_{i=1, i\ne j}^N  \int\limits_{\Omega_i} d\x \, \psi_{mn}^{-}(qr_j, \theta_j,\phi_j)
\end{align*}
(for the exterior problem, $\Omega_0 = \R^3$, and there is no second
term).  The first term can be easily computed due to orthogonality of
spherical harmonics.  In turn, one uses the II and IR re-expansion
theorems (\ref{eq:Hadd_SR}, \ref{eq:Hadd_SS}) for the second and the
third terms, respectively, in order to switch to the local spherical
coordinates of the integration domain.  After that, the corresponding
basis functions can be easily integrated.  We get
\begin{align}  \nonumber
\overline{\psi}_{mn}^j & = \frac{\sqrt{4\pi}}{q} \biggl(\delta_{n0} \delta_{m0} R_j^2 k_1(qR_j) 
- U_{mn,00}^{(-j,-0)} R_0^2 k_1(qR_0)  \\    \label{eq:int_Ij}
& - \sum\limits_{i=1, i\ne j}^N  U_{mn,00}^{(-j,+i)} R_i^2 i_1(qR_i) \biggr),
\end{align}
where $i_1(z)$ and $k_1(z)$ came from the integrals of $r^2 i_0(qr)$
and $r^2 k_0(qr)$, respectively.
Similarly, we use the RR re-expansion theorem (\ref{eq:Hadd_RR}) to get
\begin{equation}  \label{eq:int_I0}
\overline{\psi}_{mn}^0 = \frac{\sqrt{4\pi}}{q} \biggl(\delta_{n0} \delta_{m0} R_0^2 i_1(qR_0) 
- \sum\limits_{i=1}^N  U_{mn,00}^{(+0,+i)} R_i^2 i_1(qR_i) \biggr).
\end{equation}

For instance, these expressions help to find the components of the
vector $\overline{\J}$ used in Eq. (\ref{eq:tildeJ_ip}):
\begin{align}  \nonumber
\overline{\J}_{mn}^i & = \int\limits_{\Omega} d\xo \, \J_{mn}^i  \\  \nonumber
& = \frac{a_i}{R_i^2} \int\limits_{\Omega} d\xo \sum\limits_{j=0}^N \sum\limits_{k,l} \PPsi_{kl}^{j,*}(\xo)
\bigl((\q + \p\U)^{-1}  \bigr)_{kl,mn}^{ji} \\   \label{eq:tildeJ}
& = \frac{a_i}{R_i^2} \bigl(\overline{\PPsi}^* (\q + \p\U)^{-1} \bigr)_{mn}^{i} ,
\end{align}
where the vector $\overline{\PPsi}$ is formed by $\overline{\psi}_{mn}^j$.

\section{Practical implementation}
\label{sec:implementation}

A practical implementation of the GMSV requires a truncation of all
involved matrices.  If $\nmax$ denotes the truncation order for
expansions over spherical harmonics (i.e., one keeps the terms with
$n= 0,1,2\ldots,\nmax$), then the number of unknown coefficients
$A_{mn}^i$ for each $i$ is $(\nmax+1)^2$ that accounts for the second
index $m$ running from $-n$ to $n$.  In total, there are
$(N+1)(\nmax+1)^2$ unknown coefficients $A_{mn}^i$, with $i =
0,1,\ldots N$.  As discussed in detail in Ref. \cite{Grebenkov19}, the
coefficients $A_{mn}^i$ can be re-ordered to form a (row) vector as
\begin{align*}
\A = \bigl\{& \overbrace{A_{0,0}^0}^{1} , ~ \overbrace{A_{-1,0}^0,A_{0,0}^0,A_{1,0}^0}^{3} , \ldots, 
			\overbrace{A_{-n,n}^0,\ldots,A_{n,n}^0}^{2n+1}, \ldots, \\
& \, A_{0,0}^1 , ~\, A_{-1,0}^1,A_{0,0}^1,A_{1,0}^1 , \ldots, A_{-n,n}^1,\ldots,A_{n,n}^1, \ldots, \\
& ~ \ldots \hskip 14mm  \ldots  \hskip 30mm \ldots  \\
& \, A_{0,0}^N , ~\, A_{-1,0}^N,A_{0,0}^N,A_{1,0}^N , \ldots, A_{-n,n}^N,\ldots,A_{n,n}^N, \ldots\bigr\}
\end{align*}
Using the same re-ordering scheme, one can build the matrix $\W$ of
size $(N+1)(\nmax+1)^2 \times (N+1)(\nmax+1)^2$.  For the exterior
problem, there is no outer boundary $\pa_0$, all $A_{mn}^0 \equiv 0$,
and the size of the matrix is reduced to $N(\nmax+1)^2 \times
N(\nmax+1)^2$.  While larger truncation order $\nmax$ yields more
accurate results, the computational time grows very rapidly with
$\nmax$, particularly due to the matrix inversion.  When both $N$ and
$\nmax$ need to be large, the basic implementation of the GMSV is
prohibitly time-consuming, and one needs to rely on advanced
implementations (see the related discussion in
Ref. \cite{Grebenkov19}), e.g., the matrix inversion should be
implemented by iterative methods, while fast multipole methods can be
employed
\cite{Gumerov02,Gumerov05,Coifman93,Darve90,Epton95,Greengard97,Cheng06,Hesford10}.
At the same time, quite accurate results can often be achieved with
small $\nmax$ (see, e.g., Fig. \ref{fig:Jp_6balls}).

\subsection{Limits $q \to 0$ and $q\to\infty$}

In the limit $q\to 0$, the modified Helmholtz equation is reduced to
the Laplace equation, whereas the presented method becomes identical
with that from Ref. \cite{Grebenkov19}.  In particular, the basis
functions $\psi_{mn}^{\pm}$ are reduced to the basis functions
satisfying the Laplace equation:
\begin{align*}
q^{-n} \psi_{mn}^{+}(qr_i,\theta_i,\phi_i) \to \frac{\sqrt{\pi}}{2^{n+1}\Gamma(n+3/2)} \, r_i^n Y_{mn}(\theta_i,\phi_i) , \\
q^{n+1} \psi_{mn}^{-}(qr_i,\theta_i,\phi_i) \to \frac{2^n\Gamma(n+1/2)}{\sqrt{\pi}} \, r_i^{-n-1} Y_{mn}(\theta_i,\phi_i), 
\end{align*}
where we used the asymptotic behavior of the modified spherical Bessel
functions.  The re-expansion theorems and the elements of all the
matrices can thus be recalculated (see \cite{Grebenkov19} for details%
\footnote{
Since non-normalized spherical harmonics were used in Ref.
\cite{Grebenkov19}, its formulas have to be renormalized via the
normalization factor in Eq. (\ref{eq:Y}) to coincide with formulas
presented here.}).
%footnote
However, it should be noted that the singular behavior of the basis
functions $\psi_{mn}^{-}(qr_i,\theta_i,\phi_i)$ as $q\to 0$ may cause
numerical errors and instabilities for small $q$, in particular, in
the matrix inversion.  This issue should be carefully addressed upon
the implementation.

In the opposite limit of large $q$, the regular basis functions
$\psi_{mn}^{+}(qr_i,\theta_i,\phi_i)$ grow as $\exp(qr_i)$, whereas
$\psi_{mn}^{-}(qr_i,\theta_i,\phi_i)$ decays as $\exp(-qr_i)$.  This
exponential behavior may also cause numerical instabilities that can
be amended by rescaling modified spherical Bessel function by
appropriate exponential factors that can be treated explicitly.  When
the balls are well separated from each other, their diffusion
interaction is dramatically reduced in this limit, and the solution
can become much simpler.  These simplifications can be helpful for
investigating the asymptotic behavior as $q\to \infty$.

\subsection{Recurrence relations}
\label{sec:recurrence}

The direct computation of the translation matrix $\U$ via explicit
Eqs. (\ref{eq:Hadd_RR}, \ref{eq:Hadd_SR}, \ref{eq:Hadd_SS}) is
time-consuming because these formulas require numerous evaluations of
modified spherical Bessel functions, spherical harmonics, and
Clebsch-Gordan coefficients.  The computational time can be
considerably reduced by adapting the recurrence relations that were
originally derived by Chew \cite{Chew92} for the elements
$\tilde{U}_{mn,kl}^{(+j,+i)}$ of the translation operator for regular
basis functions $\tilde{\psi}_{mn}^+$ of the ordinary Helmholtz
equation (see Eq. (\ref{eq:HU_RR_ord})):
\begin{align}  \label{eq:a_beta}
& a_{mn}^+ \beta_{kl,m(n+1)} = - a_{mn}^- \beta_{kl,m(n-1)} \\  \nonumber 
& + a_{k(l-1)}^+ \beta_{k(l-1),mn} + a_{k(l+1)}^- \beta_{k(l+1),mn} , \\  \label{eq:b_beta} 
& b_{mn}^+ \beta_{kl,(m+1)(n+1)} = - b_{mn}^- \beta_{kl,(m+1)(n-1)} \\  \nonumber
& + b_{(k-1)(l-1)}^+ \beta_{(k-1)(l-1),mn} + b_{(k-1)(l+1)}^- \beta_{(k-1)(l+1),mn} ,
\end{align} 
where
\begin{align*}
a_{mn}^+ & = - \biggl(\frac{(n+1+m)(n+1-m)}{(2n+1)(2n+3)}\biggr)^{1/2} , \\
a_{mn}^- & = \biggl(\frac{(n+m)(n-m)}{(2n+1)(2n-1)}\biggr)^{1/2}  ,\\
b_{mn}^+ & = \biggl(\frac{(n+m+2)(n+m+1)}{(2n+1)(2n+3)}\biggr)^{1/2} , \\
b_{mn}^- & = \biggl(\frac{(n-m)(n-m-1)}{(2n+1)(2n-1)}\biggr)^{1/2} 
\end{align*}
for $|m|\leq n$, and $0$ otherwise \cite{Chew92} (here,
$\beta_{kl,mn}$ is a short-cut notation for
$\tilde{U}_{mn,kl}^{(+j,+i)}$, see below).  Later, Gumerov and
Duraiswami re-derived these relations and also applied them to two
other (IR and RR) re-expansion theorems \cite{Gumerov01,Gumerov02}.

\begin{figure}
\begin{center}
\includegraphics[width=80mm]{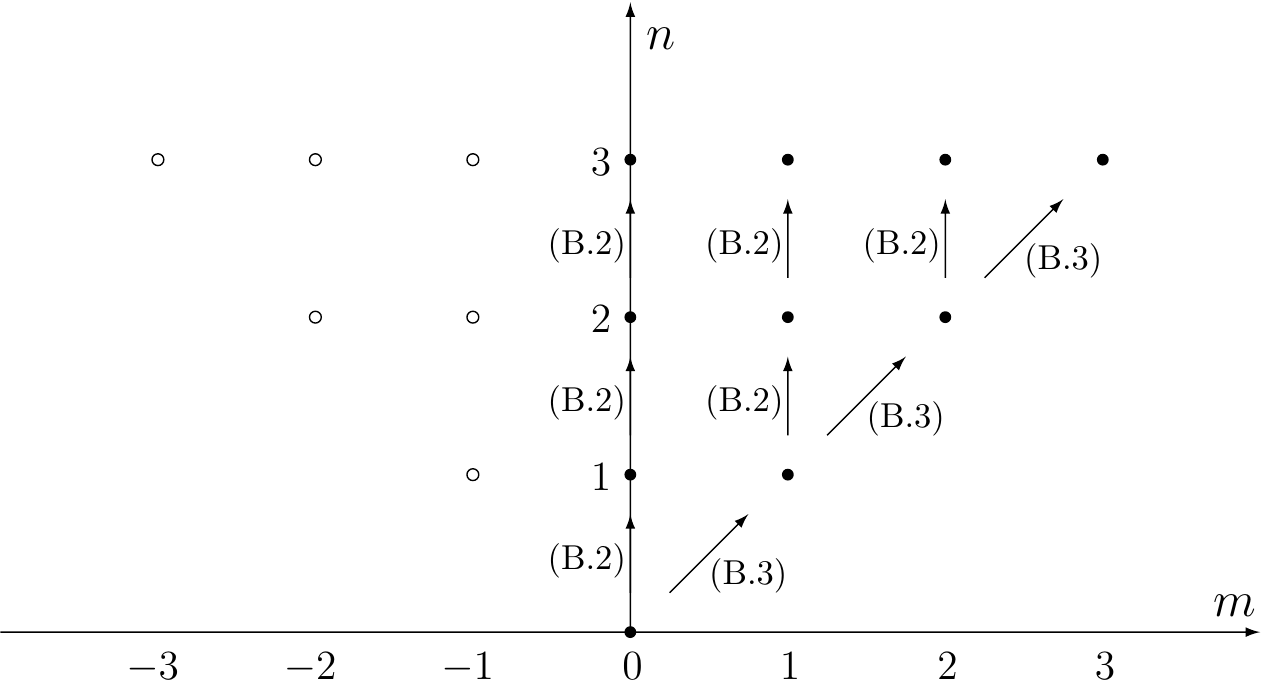} 
\end{center}
\caption{
Schematic order of the recurrence computation of $\beta_{kl,mn}$.
From the initialized values of $\beta_{kl,00}$ at $m=n=0$, one first
computes the sectorial elements $\beta_{kl,nn}$ with $n =
1,2,\ldots,\nmax$ by using Eq. (\ref{eq:b_beta}).  Then, for each $m =
1,2,\ldots,\nmax$, one moves along the $m$-th vertical line, from $n =
m$ to $n = \nmax$, using Eq. (\ref{eq:a_beta}).  In parallel, the
values for negative $m$ are computed via Eq. (\ref{eq:beta_m}).}
\label{fig:chew}
\end{figure}

As these recurrence relations result from the recurrence relations for
spherical Bessel functions and spherical harmonics, they are also
valid for the basis functions $\psi_{mn}^{\pm}$ of the modified
Helmholtz equations.  However, we could not find earlier derivations
of such recurrence relations in this setting.  Skipping tedious
mathematical details (which are similar to that presented in
\cite{Chew92,Gumerov01}), we briefly explain the use of these
relations for computing the elements of the translation matrices
$U_{mn,kl}^{(\pm j,\pm i)}$.

Let us start from the RR re-expansion theorem.  For given indices
$(k,l)$, we aim at computing recursively the elements $\beta_{kl,mn}$
for all $0\leq n \leq \nmax$ and $|m| \leq n$, where $\nmax$ is the
truncation order.  The starting point is the identity
\begin{equation}
\beta_{kl,00} = \sqrt{4\pi} \, (-1)^{k} \, i^l  \, \underbrace{i_l(qL_{ij}) \, Y_{(-k)l}(\Theta_{ij},\Phi_{ij})}
_{\psi_{(-k)l}^{+}(qL_{ij},\Theta_{ij},\Phi_{ij})} ,
\end{equation}
which follows from Eq. (\ref{eq:HU_RR}) (here we keep using the
shorter notation $\beta_{kl,mn}$ instead of $U_{mn,kl}^{(+j,+i)}$;
they slightly differ and will be related by Eq. (\ref{eq:U++_beta})).
First, one evaluates the ``sectorial'' elements $\beta_{kl,nn}$ via
the relation (\ref{eq:b_beta}).  Since $b_{nn}^{-} = 0$, the first
term on the right-hand side is canceled, and this relation expresses
$\beta_{kl,(n+1)(n+1)}$ in terms of $\beta_{k'l',nn}$ with different
indices $(k',l')$.  As a consequence, each step of the recursive
computation should be performed for the whole set of indices
$(k',l')$.  Once the sectorial elements are found, one can use the
relation (\ref{eq:a_beta}) to express $\beta_{kl,m(n+1)}$ in terms of
already known $\beta_{k'l',mn}$ and $\beta_{k'l',m(n-1)}$ (see
Fig. \ref{fig:chew}).  In this way, one can compute all the elements
up to the truncation order $\nmax$.  Note that the elements for
negative $m$ can be found from
\begin{equation}  \label{eq:beta_m}
\beta_{kl,(-m)n} = (-1)^{k+l+m+n} \beta^*_{(-k)l,mn} .
\end{equation}
We also stress that the computation of the element
$\beta_{kl,\nmax\nmax}$ via $\nmax$ repeated applications of
Eq. (\ref{eq:b_beta}) involves the element
$\beta_{(k-\nmax)(l+\nmax),00}$, so that for $l = \nmax$, one needs to
know $\beta_{(k-\nmax)(2\nmax),00}$.  As a consequence, even if the
truncation order is $\nmax$ and the translation matrix $\beta$ has to
be of the size $(\nmax+1)^2 \times (\nmax+1)^2$, intermediate
computations involve the elements of the order up to $2\nmax$.

Once the matrix elements $\beta_{kl,mn}$ are computed, one gets
\begin{equation}  \label{eq:U++_beta}
U_{mn,kl}^{(+j,+i)} = (-i)^{l-n} \beta_{kl,mn} .
\end{equation}
Similarly, one obtains the matrix elements for the II re-expansion
theorem:
\begin{equation}
U_{mn,kl}^{(-j,-i)} = i^{l-n} \beta_{kl,mn} ,
\end{equation}
which differ only by the sign factor.

Finally, in the case of the IR re-expansion theorem, the recurrence
relations are the same but they have to be initialized by using the
irregular basis function:
\begin{equation}
\tilde{\beta}_{kl,00} = \sqrt{4\pi} \, (-1)^{k+l} i^l  \, \underbrace{k_l(qL_{ij}) \, Y_{(-k)l}(\Theta_{ij},\Phi_{ij})}
_{\psi_{(-k)l}^{-}(qL_{ij},\Theta_{ij},\Phi_{ij})} 
\end{equation}
(the tilde distinguishes the matrix elements with this initialization
from the former ones).  Once such $\tilde{\beta}_{kl,mn}$ are found
using the above relations, one gets
\begin{equation}
U_{mn,kl}^{(-j,+i)} = i^{l-n} (-1)^l \tilde{\beta}_{kl,mn} .
\end{equation}

Note that modified spherical Bessel functions, their derivatives, and
spherical harmonics can also be found via standard recurrence
relations.

\subsection{Numerical inversion of the Laplace transform}
\label{sec:complex_q}

Throughout this paper, we focused on solving the modified Helmholtz
equation and thus getting solutions of time-dependent diffusion
problems in the Laplace domain.  For instance,
Sec. \ref{sec:discussion} provides semi-analytical representations for
Laplace-transformed probability flux density $\tilde{j}(\s,p|\xo)$,
first-passage time density $\tilde{H}(p|\xo)$ and reaction rate
$\tilde{J}(p)$.  Even so these quantities present their own interest,
the natural next step consists in inverting the Laplace transform to
get back to time domain.  For this purpose, one needs to compute the
Bromwich integral over a contour in the complex plane, either
numerically, or via the residue theorem.  In both cases, one has to
evaluate the quantity of interest (e.g., $\tilde{J}(p)$) at $p\in \C$,
which requires extending the presented GMSV to $q \in \C$, i.e.,
beyond the declared assumption of nonnegative $q$, see
Sec. \ref{sec:general}.  In particular, some formulas have to be
adapted to be valid for $q\in \C$.  Without pretending for generality
and rigor, we briefly discuss several lines of such extension.

Basically, one needs to check the validity of relations with complex
conjugation.  For instance, in Eq. (\ref{eq:Gf_V}), we wrote
$\PPsi_{mn}^{i,*}$ instead of $(-1)^m \PPsi_{(-m)n}^i$ which stood in
Eq. (\ref{eq:Gf}).  This identification came from Eq. (\ref{eq:Y_m})
for spherical harmonics and is valid for a real $q$, but fails for a
complex $q$.  In other words, in all relations containing $\PPsi^*$,
one has to replace $\PPsi_{mn}^{i,*}$ by $(-1)^m \PPsi_{(-m)n}^i$ to
make it valid for a complex $q$ (the same for $\overline{\PPsi}^*$).
Similarly, we employed Eq. (\ref{eq:beta_m}), which is valid for a
real $q$ but fails for a complex $q$.  For evaluating
$\beta_{kl,(-m)n}$ with a complex $q$, one can still rely on
Eq. (\ref{eq:a_beta}) with negative $m$.  In turn, the evaluation of
the sectorial element $\beta_{kl,-(n+1)(n+1)}$ can be performed by
replacing Eq. (\ref{eq:b_beta}) by
\begin{align}  
\label{eq:b_beta3} 
 b_{mn}^+ \beta_{kl,(-m-1)(n+1)} & = - b_{mn}^- \beta_{kl,(-m-1)(n-1)} \\  \nonumber
& + b_{(-k-1)(l-1)}^+ \beta_{(k+1)(l-1),(-m)n} \\  \nonumber 
& + b_{(-k-1)(l+1)}^- \beta_{(k+1)(l+1),(-m)n} .
\end{align} 
This modification allows one to evaluate the matrix elements of the
translation operators for complex $q$ and thus to apply numerical
algorithms for inverting the Laplace transform.

\section{Two concentric spheres}
\label{sec:concentric}

In this Appendix, we illustrate the use of the GMSV for a domain
between two concentric spheres of radii $R_1 < R_0$, for which the
inversion of the matrix $\W$ can be performed explicitly.  In this
domain, one has
\begin{align*}
\U^{(+0,+1)}_{mn,kl} & = \delta_{ln} \delta_{mk} \, b_{nmnm}^0 /\sqrt{4\pi}  , \\
\U^{(-1,-0)}_{mn,kl} & = \delta_{ln} \delta_{mk} \, b_{nm00}^n /\sqrt{4\pi} , 
\end{align*}
because $L_{12} = 0$ and we used $i_\nu(0) = \delta_{0\nu}$.  Since
$b_{nmnm}^0 = b_{nm00}^n = \sqrt{4\pi}$, one finds $\U^{(+0,+1)} =
\U^{(-1,-0)} = \I$ and thus
\begin{equation}
(\q + \U \p)_{mn,kl}^{ij} = \delta_{ln} \delta_{mk} v_n^{ij} ,
\end{equation}
with
\begin{subequations}  \label{eq:vn}
\begin{align}
v_n^{00} & = a_0 i_n(qR_0) + b_0 R_0 \,q\,i'_n(qR_0), \\
v_n^{01} & = a_1 i_n(qR_1) - b_1 R_1 \,q\,i'_n(qR_1), \\  
v_n^{10} & = a_0 k_n(qR_0) + b_0 R_0 \,q\, k'_n(qR_0), \\ 
v_n^{11} & = a_1 k_n(qR_1) - b_1 R_1 \,q\, k'_n(qR_1). 
\end{align}
\end{subequations}
Inverting the block diagonal matrix, we find
\begin{equation}
\bigl(\W^{-1}\bigr)_{klmn}^{ij} = \delta_{ln} \delta_{mk} \, w_n^{ij}  ,
\end{equation}
with
\begin{equation*}
\begin{split}
w_n^{00} & =  v_n^{11} /w_n ,  \qquad w_n^{01}  = - v_n^{01} / w_n,  \\
w_n^{10} & = - v_n^{10} / w_n,   \qquad w_n^{11}  = v_n^{00} / w_n,  \\
\end{split}
\end{equation*}
where
\begin{equation}
w_n = v_n^{00} v_n^{11} - v_n^{01} v_n^{10} .
\end{equation}

\subsection{Green function}

If one aims at computing the Green function $G(\x,\xo;q)$, one also
finds
\begin{subequations}
\begin{align}
F_{mn}^0 & = q \, i_n(qL_0) \, Y_{mn}^*(\Theta_0,\Phi_0) \, v_n^{10}, \\
F_{mn}^1 & = q \, k_n(qL_0) \, Y_{mn}^*(\Theta_0,\Phi_0) \, v_n^{01}, 
\end{align}
\end{subequations}
where $(L_0,\Theta_0,\Phi_0)$ are the spherical coordinates of $\xo$
(we recall that both spheres are centered at the origin).  As a
consequence, one gets the coefficients:
\begin{subequations}
\begin{align}  \nonumber
A_{mn}^0 & = q \, Y_{mn}^*(\Theta_0,\Phi_0) \\
& \times \biggl(\frac{v_n^{10} v_n^{11}}{w_n} i_n(qL_0) - \frac{v_n^{10} v_n^{01}}{w_n} k_n(qL_0) \biggr) , \\  \nonumber
A_{mn}^1 & = q \, Y_{mn}^*(\Theta_0,\Phi_0) \\
& \times \biggl(- \frac{v_n^{01} v_n^{10}}{w_n} i_n(qL_0) + \frac{v_n^{00} v_n^{01}}{w_n} k_n(qL_0)\biggr). 
\end{align}
\end{subequations}
Substituting these coefficients into Eq. (\ref{eq:gi}), we get the
Green function from Eqs. (\ref{eq:g_gi}, \ref{eq:Green0}):
\begin{align}  \nonumber
& G(\x,\xo;q) 
= G_{\rm f}(\x,\xo;q) - \frac{q}{4\pi} \sum\limits_{n=0}^\infty (2n+1) P_n\biggl(\frac{(\x \cdot \xo)}{|\x| \, |\xo|}\biggr) \\  \nonumber
& \times \biggl\{\frac{v_n^{10} v_n^{11}}{w_n} i_n(q|\xo|) - \frac{v_n^{10} v_n^{01}}{w_n} k_n(q|\xo|) \biggr) i_n(q|\x|) \\
& - \biggl(\frac{v_n^{01} v_n^{10}}{w_n} i_n(q|\xo|) - \frac{v_n^{00} v_n^{01}}{w_n} k_n(q|\xo|)\biggr) k_n(q|\x|)  \biggr\} ,
\end{align}
where $P_n(z)$ are Legendre polynomials, and we used the addition
theorem for spherical harmonics to perform the sum over $m$:
\begin{equation}
\sum\limits_{m=-n}^n Y_{mn}(\theta,\phi) Y_{mn}^*(\Theta_0,\Phi_0) = \frac{2n+1}{4\pi} 
P_n\biggl(\frac{(\x \cdot \xo)}{|\x| \, |\xo|}\biggr).
\end{equation}

This general expression is reduced to two limiting cases:

(i) an interior problem inside a sphere of radius $R_0$ corresponds to
the limit $R_1 \to 0$, in which $v_n^{01} \to 0$ while $v_n^{11} \to
\infty$ so that
\begin{align} \nonumber
& G(\x,\xo;q) = G_{\rm f}(\x,\xo;q) - \frac{q}{4\pi} \sum\limits_{n=0}^\infty (2n+1) P_n\biggl(\frac{(\x \cdot \xo)}{|\x| \, |\xo|}\biggr) \\ 
\label{eq:GH_oneD} 
& \times \frac{a_0 k_n(qR_0) + b_0 R_0 q k'_n(qR_0)}{a_0 i_n(qR_0) + b_0 R_0 q i'_n(qR_0)} \, i_n(q|\x|) \, i_n(q|\xo|) ;
\end{align}

(ii) an exterior problem outside one sphere of radius $R_1$
corresponds to the limit $R_0 \to\infty$, in which $v_n^{10} \to 0$
while $v_n^{00} \to \infty$ so that
\begin{align}  \nonumber
& G(\x,\xo;q) = G_{\rm f}(\x,\xo;q) - \frac{q}{4\pi} \sum\limits_{n=0}^\infty (2n+1) P_n\biggl(\frac{(\x \cdot \xo)}{|\x| \, |\xo|}\biggr) \\ 
\label{eq:GH_oneE}
& \times \frac{a_1 i_n(qR_1) - b_1 R_1 q i'_n(qR_1)}{a_1 k_n(qR_1) - b_1 R_1 qk'_n(qR_1)} \, k_n(q|\x|) \, k_n(q|\xo|) . 
\end{align}

Note that the distribution of the reaction time for two concentric
spheres was studied in Ref. \cite{Grebenkov18c}.

\subsection{Dirichlet-to-Neumann operator}

Using the above explicit relations, we also compute the matrix $\MM$
determining the spectrum of the Dirichlet-to-Neumann operator:
\begin{align}  
& (\MM)_{mn,kl} = \delta_{nl} \delta_{mk} \frac{q}{w_n}  \times  \\  \nonumber  
& \small \biggl(\begin{array}{ll}
v_n^{11} i'_n(qR_0) - v_n^{01} k'_n(qR_0) & -v_n^{11} i'_n(qR_1) + v_n^{01} k'_n(qR_1) \\
-v_n^{10} i'_n(qR_0) + v_n^{00} k'_n(qR_0) & v_n^{10} i'_n(qR_1) - v_n^{00} k'_n(qR_1) \\  \end{array} \biggr).
\end{align}

Let us first consider the case when only the inner sphere is reactive
whereas the outer sphere is reflecting.  Substituting $a_0 = 0$, $a_1
= 1$, $b_0 = 1$, $b_1 = 0$ into Eqs. (\ref{eq:vn}), we get
\begin{align}  
& (\MM)_{mn,kl} = \delta_{nl} \delta_{mk}  \\  \nonumber
& \times \left(\begin{array}{cc}
1/R_0 & -1/(qR_1)^2/R_0 \\
0 & q \frac{k'_n(qR_0) i'_n(qR_1) - i'_n(qR_0) k'_n(qR_1)}{k_n(qR_1) i'_n(qR_0) - i_n(qR_1) k'_n(qR_0)} \\  \end{array} \right).
\end{align}
The diagonal structure of this matrix allows one to easily determine
its eigenvalues:
\begin{equation}
\mu_n^{(p)} = q\frac{k'_n(qR_0) i'_n(qR_1) - i'_n(qR_0) k'_n(qR_1)}{k_n(qR_1) i'_n(qR_0) - i_n(qR_1) k'_n(qR_0)}  \,,
\end{equation}
where $q = \sqrt{p/D}$.  Note that this matrix also has infinitely
many spurious eigenvalues $1/R_0$, which come from the redundant form
of the matrix $\MM$ in this setting with Neumann condition.
Similarly, one can treat the case when only the outer sphere is
reactive.

When both spheres are reactive, one substitutes $a_0 = a_1 = 1$ and
$b_0 = b_1 = 0$ into Eqs. (\ref{eq:vn}) to get
\begin{widetext}
\begin{align}  
& (\MM)_{mn,kl} =  \frac{\delta_{nl} \delta_{mk} \, q}{i_n(qR_0) k_n(qR_1) - i_n(qR_1) k_n(qR_0)} \\  \nonumber
& \times \left(\begin{array}{cc}
k_n(qR_1) i'_n(qR_0) - i_n(qR_1) k'_n(qR_0) & -1/(qR_1)^2 \\
-1/(qR_0)^2 & k_n(qR_0) i'_n(qR_1) - i_n(qR_0) k'_n(qR_1) \\  \end{array} \right).
\end{align}
\end{widetext}
The eigenvalues are obtained by diagonalizing separately each $2\times
2$ block of this matrix and can be expressed as solutions of the
associated quadratic equation.

%%%%%%%%%%%%%%%%%%%%%%%%%%%%%%%%%%%%%%%%%%%%%%%%%%%%%%%%%%%%%%%%%%%%%%%%%%%%%%%%%%%%%%

\end{document}